\documentclass[11pt,a4paper]{article}
\pdfoutput=1
\usepackage{jcappub}
\usepackage{aas_macros_JCAP}
\usepackage{blindtext}
\usepackage{bm}
\usepackage{verbatim}

\newcommand{\Msun}{M_{\odot}/h}
\newcommand{\Mpch}{{\textrm{Mpc}}/h}
\newcommand{\void}{\mathrm{v}}
\newcommand{\halo}{\mathrm{h}}

\newcommand{\tracer}{\mathrm{t}}
\newcommand{\matter}{\mathrm{m}}

\newcommand{\coreDens}{\hat{n}_\mathrm{min}}

\newcommand{\hydro}{\texttt{hydro}}
\newcommand{\DMo}{\texttt{DMo}}

\newcommand{\mr}{\texttt{mr}}

\newcommand{\hr}{\texttt{hr}}
\newcommand{\MR}{\texttt{midres}}
\newcommand{\HR}{\texttt{highres}}
\newcommand{\UHR}{\texttt{ultra-hr}}
\newcommand{\uhr}{\texttt{uhr}}
\newcommand{\Mag}{\texttt{Magneticum}}

\title{Why Cosmic Voids Matter: Pristine~Evolution}

\author[a,b]{Nico Schuster,}
\author[b]{Nico Hamaus,}
\author[a,c ]{Alice Pisani,}
\author[b,d]{Klaus Dolag,}
\author[b,e,f]{Jochen Weller}

\affiliation[a]{Aix-Marseille Universit\'e, CNRS/IN2P3, CPPM, Marseille, France}
\affiliation[b]{Universit\"ats-Sternwarte M\"unchen, Fakult\"at f\"ur Physik, Ludwig-Maximilians-Universit\"at, Scheinerstr. 1, 81679 M\"unchen, Germany}
\affiliation[c]{Department of Astrophysical Sciences, Peyton Hall, Princeton University, Princeton, NJ 08544, USA}
\affiliation[d]{Max-Planck-Institut f\"ur Astrophysik, Karl-Schwarzschild-Str. 1, 85748 Garching, Germany }
\affiliation[e]{Excellence Cluster ORIGINS, Bolzmannstr. 2, 85748 Garching, Germany}
\affiliation[f]{Max-Planck-Institut f\"ur extraterrestrische Physik, Giessenbachstr. 1, 85748 Garching, Germany}

\emailAdd{schuster@cppm.in2p3.fr}
\emailAdd{n.hamaus@physik.lmu.de}
\emailAdd{pisani@cppm.in2p3.fr}
\emailAdd{kdolag@mpa-garching.mpg.de}
\emailAdd{jochen.weller@lmu.de}

\abstract{We utilize the Magneticum suite of hydrodynamical simulations to investigate the formation and evolution of cosmic voids from $z = 5.04$ to present day, using cold dark matter and (sub-) halo tracers in high-density samples. This includes the evolution of their global properties, such as size, shape, inner density, and average density, as well as their radial density profiles. Our results provide several key conclusions for void analyses in modern surveys. We demonstrate that a relative size framework is required, mitigating methodological selection effects and revealing the true physical evolution of densities around halo-defined voids. This necessity arises from our findings that a void’s properties are more fundamentally tied to its rank within its contemporary population than to its absolute size. Using this framework, we show that the evolution of halo voids stabilizes at redshifts below $z \simeq 1$, driven primarily by cosmic expansion rather than ongoing halo formation. We further find that the matter evolution around these stable voids is remarkably well-described by linear growth theory, with deviations appearing as non-linear growth on small scales and suppressed growth in the largest voids, potentially driven by the influence of dark energy. This late-time stability and the predictable evolution confirm voids as pristine laboratories for probing the nature of dark energy with upcoming surveys.}

\date{\today}


\begin{document}
\maketitle

\newpage

\section{Introduction\label{sec:intro}}

The cosmic web that we observe today is a vast network consisting of clusters, filaments, and cosmic voids~\cite{Gregory1978, Joeveer1978, Kirshner1981, Zeldovich1982, Bertschinger1985, vdWeygaert1993}. While voids are now the largest and most underdense structures in our Universe, they grew from tiny initial density fluctuations whose imprints are still visible in the cosmic microwave background (CMB)~\cite{Bennett2003, Planck2020}. These primordial seeds evolved under the influence of gravity and cosmic expansion to form the large-scale structures seen today. The prevailing cosmological model, known as the $\Lambda$CDM model, assumes that these structures are primarily composed of a form of matter that interacts only gravitationally, known as \textit{cold dark matter} (CDM). The late-time accelerated expansion of the Universe is, in turn, attributed to an unknown \textit{dark energy}, which is so far consistent with a cosmological constant, $\Lambda$~\cite{Riess1998, Perlmutter1999}, although recent observations have begun to show potential hints for a more dynamic nature of dark energy~\cite{DESI_DR1_constraints_bao, DESI_DR1_constraints_full_shape, DESI_DR2_constraints_bao}.

The study of cosmic voids has evolved significantly since their initial discovery in early galaxy surveys of the 1970s~\cite{Gregory1978, Joeveer1978}. The advent of modern cosmological surveys, such as SDSS~\cite{SDSS}, BOSS~\cite{Dawson2013}, DES~\cite{DarkEnergySurvey2005}, and eBOSS~\cite{Dawson2016}, has since established them as both essential laboratories for modern cosmology~\cite{Pisani2019, Moresco2022, DiValentino2025, Cai2025} and powerful probes of large-scale structure~\cite{Biswas2010, Pan2012, Lavaux2012, Sutter2012b, Hamaus2014a, Pisani2015a, Mao2017b, Hamaus2017,Sahlen2019, Nadathur2020b, Hamaus2020, Correa2021a, Davies2021,Vielzeuf2021,Aubert2022, Kovacs2022}. Among the most commonly studied statistics is the abundance of voids by their size, known as the void size function (VSF)~\cite{Sheth2004,Conroy2005, Jennings2013, Pisani2015a, Nadathur2016, Sahlen2016, Contarini2019, Verza2019, Contarini2023, Verza2024b}. Additionally, two-point functions such as the void-galaxy cross-correlation function~\cite{Hamaus2014b, Pisani2014, Cai2016, Hamaus2016, Hawken2017, Achitouv2019, Correa2019, Hawken2020, Correa2021b, Hamaus2022, Woodfinden2022, Radinovic2023}, void lensing~\cite{Krause2013, Clampitt2015, Chantavat2017, Cai2017, SanchezC2017, Baker2018, Fang2019, Bonici2023,Martin2025} and the void auto-correlation~\cite{Hamaus2014a,Hamaus2014c, Chan2014, Chuang2017, Kreisch2019, Voivodic2020, Kreisch2022} are remarkable tools to constrain cosmology and structure formation. Most theoretical models for these statistics begin by predicting voids in the underlying distribution of matter and then linking them to voids found in biased tracers like galaxies. More recently, modern machine learning tools have also been employed to investigate voids and their statistics~\cite{Kreisch2022, Wang2023, Thiele2024,Wang2024, Fraser2025, Lehman2025, Salcedo2025}.

The power of cosmic voids as cosmological probes is rooted in their unique physical characteristics. Their interiors, unlike collapsed structures, remain coupled to the Hubble flow, and the low density contrast within them ensures the dynamics of matter are well-described by linear theory down to the smallest cosmological scales~\cite{Hamaus2014b, Schuster2023}. Furthermore, voids are largely insensitive to complex baryonic physics~\cite{Schuster2024}, making them pristine laboratories governed almost exclusively by gravity. Combined, these characteristics not only allow for simpler and more robust modeling compared to dense environments~\cite{Stopyra2021, Lepinzan2025}, but also make void statistics highly sensitive to fundamental physics, such as the effects of massive neutrinos~\cite[e.g.,][]{Massara2015, Banerjee2016,Kreisch2019, Schuster2019, Zhang2020, Contarini2021, Bayer2021, Kreisch2022, Thiele2024}, modifications to General Relativity(GR)~\cite[e.g.,][]{Zivick2015, Cai2015, Barreira2015, Falck2018, Baker2018, Paillas2019, Davies2019, Perico2019, Wilson2021, Tamosiunas2022, Fiorini2022, Williams2025}, the exploration of the nature of dark matter~\cite[e.g.,][]{Yang2015, Reed2015, Baldi2018, Lester2021, Arcari2022, Lester2025}, and inflation~\cite{Chan2018}.

To date, the study of cosmic voids has predominantly focused on the low-redshift Universe, with comparatively few studies exploring their full evolutionary history through theoretical models~\cite[e.g.,][]{vdWeygaert2016, Gallagher2022, Pizana2024, Bromley2025} and cosmological simulations~\cite[e.g.,][]{Ceccarelli2013, Paz2013, Ruiz2015, Cautun2014,Sutter2014c, Wojtak2016,Adermann2018, Massara2018,VallesPerez2021, Verza2023, Curtis2025}. The advent of current and upcoming spectroscopic galaxy surveys such as DESI~\cite{DESI2025_DR1, Rincon2025}, Euclid~\cite{Hamaus2022, Contarini2022, Bonici2023, Radinovic2023}, Roman~\cite{Verza2024b}, and SPHEREx~\cite{Dore2018}, is opening a new observational frontier, enabling the study of voids in dense samples across wide redshift ranges.

A robust understanding of void evolution is therefore essential to fully leverage these forthcoming datasets, providing the primary motivation for the detailed investigation presented in this paper. Building upon previous studies in this series that investigated watershed voids at fixed low redshifts in hydrodynamical simulations~\cite{Schuster2023, Schuster2024}, we now examine the evolution of their statistical properties from redshift $ z = 5.04$ to present day, covering over $12.5 \, \mathrm{Gyr}$ of cosmic history across a variety of simulation resolutions and volumes.

We begin by investigating the evolution of the underlying tracer distributions (CDM and halos) and the global properties of the voids identified within them, such as their size, shape, and average density. Our focus then shifts to a detailed study of their internal structure using density profiles. For halo-defined voids (hereafter, halo voids), we examine both the halo number density and the surrounding CDM distribution, a configuration highly relevant for weak lensing studies. These CDM profiles additionally provide a framework for a comparison with theoretical predictions of structure growth. To ensure a clear comparison between these results, each redshift corresponds to a specific color, linestyle, and marker throughout.

The paper is organized as follows. Section~\ref{sec:Magneticum} describes the \Mag{} simulation suite used in this work, while Section~\ref{sec:methods} details our void-finding and profile-estimation methods. In Section~\ref{sec:catalogs}, we characterize the evolution of the tracer populations and the global properties of the voids identified within them. Section~\ref{sec:density_profiles} presents our detailed analysis of the evolution of void density profiles and the impact of void selection on this evolution. In Section~\ref{sec:growth_factor}, we test the predictions of linear structure growth against our measured profiles for both evolving and fixed void populations. Finally, we summarize our findings and conclude in Section~\ref{sec:conclusion}.

\section{The Magneticum simulations}
\label{sec:Magneticum}

\begin{table}[htbp]
\centering
\begin{tabular}{|c | c c c c c c|} 
 \hline
 Name & {\it Box} & $L_\mathrm{Box}$ & $N_\mathrm{particles}$ & $m_\mathrm{CDM}$ & $m_\mathrm{baryon}$ & $z_\mathrm{min}$ \\ [0.5ex] 
 \hline
 \rule{0pt}{3ex}
 \MR{} (\mr{}) & {\it 0} & 2688 & $2 \times 4536^3$ & $1.3 \times 10^{10}$ & $2.6 \times 10^9$ & 0.00  \\
 \HR{} (\hr{}) & {\it 2b} & 640 & $ 2 \times 2880^3$ & $6.9 \times 10^8$ & $1.4 \times 10^8$ & 0.25 \\
 \UHR{} (\uhr{}) & {\it 4} & 48 & $ 2 \times 576^3$ & $3.6 \times 10^7$ & $7.3 \times 10^6$ & 0.07 \\ [1ex]
 \hline
\end{tabular}
\caption{Properties of the \Mag{} simulations analyzed in this work. The box length $L_\mathrm{Box}$ is in units of $\Mpch$ and particle masses ($m_\mathrm{CDM}$ \& $m_\mathrm{baryon}$) are in units of $\Msun$.}
\label{table_1}
\end{table}

This paper analyzes voids using the \Mag\footnote{\url{http://www.magneticum.org}} suite of state-of-the-art cosmological hydrodynamical (\hydro{}) and dark-matter-only (\DMo{}) simulations, which cover a wide range of cosmological volumes and mass resolutions. We exclusively analyze the \hydro{} runs, as baryonic effects on voids have already been investigated in previous work of this series~\cite{Schuster2024}. For clarity, we hereafter use the term `simulations' to refer to the \hydro{} runs throughout the remainder of this paper.

We provide a brief overview of the \Mag{} suite here and refer the reader to previous works for more details~\cite[e.g.,][]{Hirschmann2014,Dolag2015, Steinborn2015,Teklu2015,Bocquet2016,Dolag2016,Remus2017,Castro2018,Castro2021,Angelinelli2022}, and in particular to Reference~\cite{Dolag2025} for a comprehensive description. All the simulations analyzed in this work adopt a flat $\Lambda$CDM cosmology consistent with the best-fit parameters of WMAP7 (see~\cite{Komatsu2011}), $\Omega_{\Lambda} = 0.728$, $\Omega_\mathrm{m} = 0.272$, $\Omega_\mathrm{b} = 0.0456$, $h = 0.704$, $\sigma_8 = 0.809$, and $n_s = 0.963$. They were performed using an advanced version of the TreePM-SPH code \textsc{gadget3}~\cite{Springel2005}, which incorporates an improved SPH solver~\cite{Beck2016}. In addition, the code implements a comprehensive model for baryonic physics, including chemical enrichment from multiple metal species~\cite{Dolag2017}, as well as models for black hole growth and active galactic nuclei (AGN) feedback based on previous work~\cite{Springel2005a,DiMatteo2005,Fabjan2010}. A summary of the key parameters for the \hydro{} runs, such as box sizes, particle counts, and mass resolution for both baryons and CDM particles, is provided in Table~\ref{table_1}.

To properly trace the evolution of large-scale structures, our analysis primarily focuses on Box0 and Box2b. These simulations contain a sufficient number of cosmic voids in a large volume across multiple redshifts, allowing us to investigate the statistical evolution of voids and their relevant scales across time. Previous work using these simulations demonstrates their versatility for the study of voids~\cite{Pollina2017,Pollina2019,Schuster2023, Pelliciari2023, Schuster2024,Lehman2025}. For simplicity, we will refer to these simulations by their respective mass resolutions throughout this paper, i.e. \MR{} (\mr{}) for {\it Box0} and \HR{} (\hr{}) for {\it Box2b}. To test the scale-dependence of theoretical predictions, we also include {\it Box4} with its ultra-high resolution, although it contains significantly fewer voids. This box will be referred to as \UHR{} (\uhr{}).

For the subsequent analysis of voids in biased tracers, we identify subhalos and their properties using the \textsc{SubFind} algorithm \cite{Springel2001b}, which has been modified to account for baryonic components \cite{Dolag2009}. For simplicity, we will refer to subhalos as `halos' from this point forward. We have verified that all results presented in this paper are robust to this substitution and do not change qualitatively when using the actual halo sample instead of subhalos.

\section{Methodology \label{sec:methods}}

\subsection{VIDE void finding \label{subsec:void_finding}}

To identify voids in the distributions of halos and the underlying CDM, we use the Void IDentification and Examination toolkit {\textsc{vide}}\footnote{\url{https://bitbucket.org/cosmicvoids/vide_public/}}~\cite{Sutter2015}, which is based on the ZOnes Bordering On Voidness (\textsc{zobov}) algorithm~\cite{Neyrinck2008}. This watershed algorithm (see~\cite{Platen2007}) identifies local underdensities in the three-dimensional density field constructed via Voronoi tessellation. For each tracer particle, \textit{j}, a unique Voronoi cell with volume $\textit{V}_j$ is assigned as the region of space closest to that tracer. The density within this cell is defined as the inverse of its volume ($\rho_j = 1 / \textit{V}_j$). The watershed algorithm then searches for extended density depressions by starting at the local density minima and monotonically increasing the density until it drops again, thereby defining the underdense cores and boundaries of voids.

The merging of adjacent voids is controlled by a density-based threshold, which is a free and optional parameter in \textsc{vide}. It is important to note that this is completely unrelated to the merging of cosmic voids through their evolution across time. The threshold determines whether adjacent voids are merged: two voids are added together if the density on their shared wall is below the threshold multiplied by the mean number density of tracers, $\bar{n}_\tracer$. This produces a hierarchy of parent and child voids. However, we adopt the very low default threshold of $10^{-9}$, which effectively prevents all merging and ensures our void catalogs consist exclusively of \emph{isolated} voids. Consequently, we will not explicitly use the term \emph{isolated} voids for the remainder of this paper. For a more comprehensive discussion on void merging and its effect on void statistics, we refer to previous work from this series \cite{Schuster2023}.

Because \textsc{vide} identifies voids without making any assumptions about their shapes, the resulting catalogs consist of non-spherical voids with a variety of properties. The center of a particular void is defined as the volume-weighted barycenter of all its member particles:

\begin{equation}
\label{eq:void_barycenter}
\mathbf{X}_\mathrm{v} = \frac{\sum_j \, \bm{x}_j \, \textit{V}_j}{\sum_j \, \textit{V}_j} \,,
\end{equation}

where $\bm{x}_j$ are the comoving tracer positions. This barycenter is not necessarily the location of the lowest-density cell, but rather represents the geometric center of the void, primarily constrained by its boundary. This definition makes the center robust against Poisson fluctuations in tracer positions and representative of the whole void structure.

Given their non-spherical nature, voids can only be characterized by an \emph{effective void radius} $r_\void$, defined as the radius of a sphere with identical volume as the void. This volume is calculated by summing the volumes of all the associated Voronoi cells of a void:

\begin{equation}
\label{eq:void_radius}
r_\void = \left( \frac{3}{4 \pi} \sum_j  \, \textit{V}_j \right)^{1/3} .
\end{equation}

To quantify the non-spherical shape of voids, \textsc{vide} calculates the ellipticity using the largest ($J_3$) and smallest ($J_1$) eigenvalues of the inertia tensor (see~\cite{Sutter2015,Schuster2023} for additional details). The ellipticity is then given by:

\begin{equation}
\label{eq:void_ellipticity}
\varepsilon = 1 - \left( \frac{J_1}{J_3} \right)^{1/4} \,.
\end{equation}

We also investigate the compensation $\Delta_\tracer$ and core density $\coreDens$ as additional void properties. The compensation is a measure of a void's local environment, indicating whether it contains more or fewer member particles ($N_\tracer$) than an average region of the Universe with the same volume ($V$). Voids with $\Delta_\tracer < 0$ are classified as undercompensated, and those with $\Delta_\tracer > 0$ are overcompensated~\cite{Hamaus2014a,Hamaus2014b}. This compensation is defined as:

\begin{equation}
\label{eq:void_compensation}
\Delta_\tracer \equiv \frac{N_\tracer/V}{\bar{n}} - 1 = \hat{n}_{\mathrm{avg}} - 1\,.
\end{equation}

Lastly, the core density is defined as the minimal density within a void, which is given by the density of its largest Voronoi cell and is expressed in units of the mean tracer density:

\begin{equation}
\label{eq:void_coreDensity}
\coreDens =   \frac{ n_{\mathrm{min}}   }{  \bar{n} }\,.
\end{equation}

\subsection{Void profiles \label{subsec:void_profiles}}

\begin{table}[htbp]
\centering
\begin{tabular}{|c  | c  |} 
 \hline
 Type of profile &  Formula \\
 \hline
  \rule{0pt}{4.5ex} 
number density, individual & $n_\void^{(i)}(r) = \frac{3}{4\pi \, } \sum_{j } \frac{ \Theta(r_j)}{ \left( r + \delta r \right)^3 - \left( r - \delta r \right)^3 }$  \\
 \rule{0pt}{4.5ex} 
number density, stacked & $n_\void(r) = \frac{1}{N_\void} \sum\limits_{i} n_\void^{(i)}(r)$   \\ [2.5ex] 
 
 \hline
\end{tabular}
\caption{Formulas for calculating the void number profiles presented in this work. For more details, a discussion on weights in the density profiles, as well as a detailed description of void velocity profiles, we refer the reader to sections 3.2 and 5.3 in Reference~\cite{Schuster2023}. The meaning of the variables is explained in the text. The code base for the estimation of these profiles is available {here}\footref{note:data_url}.}
\label{table_2}
\end{table}

The full code base for estimating density and velocity profiles, including the various stacking methods, is publicly available {here}\footnote{\label{note:data_url}\url{https://github.com/nicosmo/void_profile_analysis}}. For a more detailed description of density and velocity profiles and their different estimators, we refer to Section 3.2 of Reference \cite{Schuster2023}.

We present a summary of the formulas used to calculate void number density profiles in Table~\ref{table_2}, which are the focus of our subsequent redshift evolution analysis. The profiles represent the spherically averaged density contrast from the mean value, $\bar{n}$, and are calculated in spherical shells around void centers. For each individual void, we calculate these profiles using a constant radial bin size expressed as $\delta r/r_\void = \mathrm{const}$. This ensures that characteristic features, such as the compensation walls near $r = r_\void$, are located at the same relative position from void centers. This allows consistent comparisons between individual and stacked profiles for voids of different sizes.

In the formula for the individual density profiles in table~\ref{table_2}, the function $\Theta( r_j)$ combines two Heaviside step functions $\vartheta$ to define the radial bins of the profile. It is defined as $\Theta( r_j) \equiv \vartheta \left[ r_j - \left( r - \delta r \right) \right] \, \vartheta \left[- r_j + \left( r + \delta r \right) \right] $, with $r_j$ being the distance of tracer $j$ from the center of the void. The summation over $j$ includes all tracers within a specified vicinity of the voids. Stacked profiles are then calculated by averaging individual void profiles, typically grouped by void size. The errors on these stacked profiles are estimated using the jackknife method.

It is important to note that when plotting halo number densities, we use the contrast $ n / \bar{n} -1$. However, for CDM tracers, the number density is identical to the actual density, so we use the contrast $\rho / \bar{\rho} -1 $.

\section{Tracer and void properties \label{sec:catalogs}}

\subsection{Halo and void population evolution \label{subsec:halo_population}}

Before analyzing the evolution of common void properties, we must first examine the evolution of the halo population itself. The number of halos above specific mass cuts directly determines both the number of voids identified therein and the physical scales probed by the analysis. We select these mass cuts to yield a sample of reliable, well-resolved tracers for void identification.

\begin{figure}[t!]
               \centering
               \resizebox{\hsize}{!}{
                               \includegraphics[trim=7 7 0 7, clip]{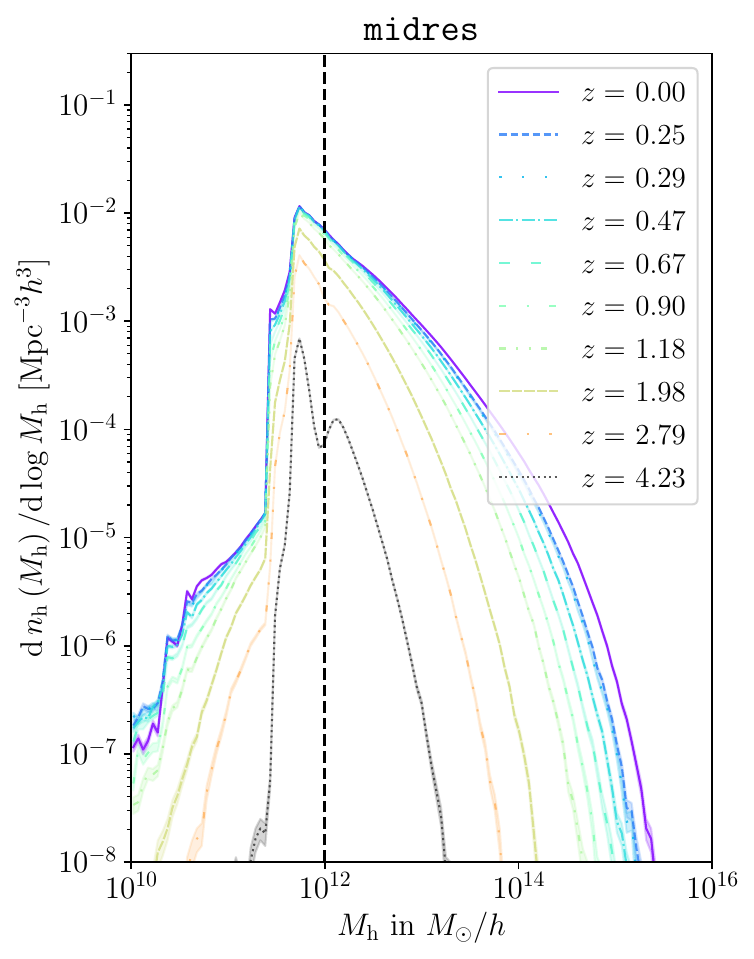}
                               
                               \includegraphics[trim=7 7 7 7, clip]{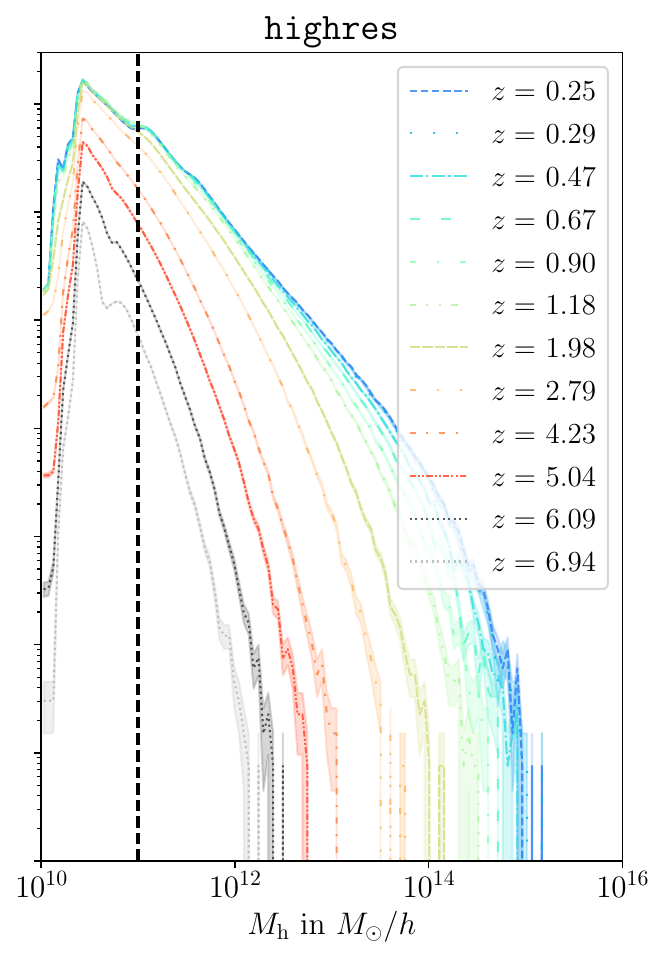}}
                               
               \caption{Redshift evolution of the halo mass function (HMF) in \MR{} (left) and \HR{} (right), with redshifts given in the legends. The vertical lines indicate the halo mass cuts of $1.0 \times 10^{12} \, \Msun $ (\mr{}) and $1.0 \times 10^{11} \, \Msun $ (\hr{}) that are chosen to select halos for the void identification at redshifts $z \leq 2.79$ in \mr{} and $z \leq 5.04$ in \hr{}.}
               \label{fig_halo_mass_functions}
\end{figure}

\begin{figure}[t!]
               \centering
               \resizebox{\hsize}{!}{
                               \includegraphics[trim=7 7 0 7, clip]{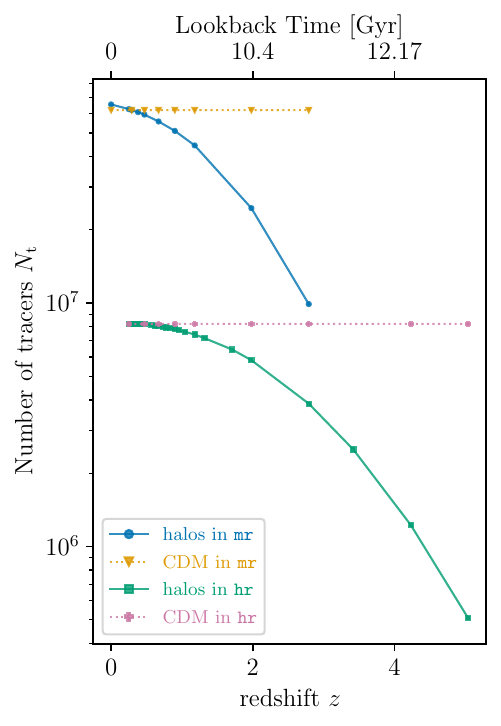}
                               
                               \includegraphics[trim=7 7 7 7, clip]{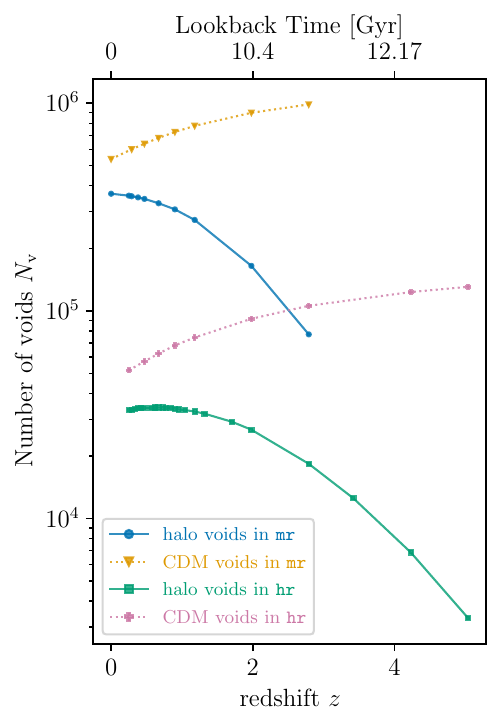}}

               \caption{Redshift evolution of tracer numbers (left) and number of voids (right) identified therein. Halos (solid lines) included in the void identification are selected with $M_\halo \geq 10^{12} \, \Msun$ in \MR{}, while in \HR{} we utilize $M_\halo \geq 10^{11} \, \Msun $. At all redshifts, CDM voids are identified in fixed CDM subsampling (dashed lines) of approximately $62 \times 10^6$ particles in \mr{} and  $8.2 \times 10^6$ in \hr{}. This figure reveals the divergent evolution of the two void populations: CDM voids merge and decrease in number, while halo voids fragment and increase in number as more halos form over time.}
               \label{fig_tracer_void_numbers}
\end{figure}

Figure~\ref{fig_halo_mass_functions} depicts the evolution of the (sub-) halo mass function (HMF) across a range of redshifts in the \MR{} and \HR{} simulations. The vertical lines mark the selected cuts in halo mass: $1.0 \times 10^{12} \, \Msun$ in \mr{} and $1.0 \times 10^{11} \, \Msun$ in \hr{}. These constant mass cuts ensure that the impact of halo formation and evolution is included in the corresponding evolution of halo voids. As expected, the number of halos above the mass cuts steadily increases, and they become more massive as the redshift decreases. Starting at $z = 1.98$, the HMF in \hr{} becomes relatively stationary close to $M_\halo \lesssim  1.0 \times 10^{11} \, \Msun$, with continued growth observed only at higher masses. This suggests an approximate equilibrium between the formation of new halos and the growth of existing ones through mergers and matter accretion.

Due to the lower resolution in \mr{}, the selected mass cut is $1.0 \times 10^{12} \, \Msun$, yet the HMF exhibits an identical evolution to halos with $M_\halo \geq 1.0 \times 10^{11} \, \Msun$ in \hr{}. However, at $z = 4.23$, the HMF in \mr{} experiences a slight dip near this chosen mass cut. Given this dip, we expect that at this redshift the number of halos above the mass cut is smaller than it should be. This would consequently lead to fewer voids being identified, directly influencing the void size function and potentially other void statistics. We therefore exclude $z \geq 4.23$ in our further analysis of voids in \mr{}. In contrast, the HMF in \hr{} only includes a dip for $z = 6.94$, located at masses below $1.0 \times 10^{11} \, \Msun$, allowing us to place greater confidence in void statistics at these high redshifts. Even so, the number of voids at redshifts $z > 5.04$ constitutes only $\approx 3 \%$ or less of the void catalogs at $ z \lesssim 1 $. Hence, we also refrain from including these in further analyses, instead focusing on voids identified at $z \leq 5.04$.

Figure~\ref{fig_tracer_void_numbers} displays the evolving numbers of halos, halo voids, and voids identified in constant CDM subsamplings. These CDM subsamplings are chosen to match the halo number densities at low redshift ($z = 0.29$, see~\cite{Schuster2023}), with CDM tracer numbers of $62 \times 10^6$ ($0.066\%$ of all CDM particles in \mr{}) and $8.2 \times 10^6$ ($0.034\%$ in \hr{}). This approach enables the identification of voids of comparable scales in both halos and CDM at low redshift, while allowing us to cleanly separate the physical evolution of CDM from the effects of halo formation. Continuously adjusting the CDM density to match the halo number density at each redshift would instead entangle halo formation effects with CDM void statistics. This choice of subsampling is similar to smoothing the density field near scales of the mean tracer separation without imposing an explicit smoothing length.

For the \MR{} simulation, Figure~\ref{fig_tracer_void_numbers} presents data from $z = 0.00$ to $z = 2.79$, reflecting the resolution limit of the HMF at higher redshifts, as discussed in Figure~\ref{fig_halo_mass_functions}. In the \HR{} simulation,  counts are displayed from $z = 5.04$ down to $z = 0.25$, which is when the simulation ended. We exclude data at higher redshifts ($z > 5.04$) from this figure because the low abundance of resolved halos yields a void sample too small to be representative of the underlying cosmic web.

Additionally, halo voids are analyzed at more redshifts than CDM voids. While not all of these redshifts appear in plots of void property distributions or profiles, the omitted void statistics consistently follow the overall evolutionary trends observed between the included redshifts.

In fixed CDM subsampling for both \mr{} and \hr{}, the number of CDM voids decreases with decreasing redshift. At the highest redshifts in Figure~\ref{fig_tracer_void_numbers}, we identify approximately $9.9 \times 10^5$ CDM voids in \mr{} and $1.3 \times 10^5$ in \hr{}, which drop to around $5.4 \times 10^5$ and $0.52 \times 10^5$, respectively, at the lowest redshifts. This decline reflects the transition from a more uniform CDM distribution, characterized by numerous small underdensities in the early Universe, to more prominent filamentary structures in the cosmic web. Many smaller voids merge into larger ones as their separating structures collapse and are attracted to density peaks, causing some void boundaries to disappear.

In contrast, Reference~\cite{Sutter2014c} reports an increase in CDM void numbers over time. Their analysis applied a lower limit on void size and used the full CDM particle distribution for identification, which tends to yield smaller voids. In our work, we do not impose a minimum void size and include fewer particles; we thus attribute the differing trends in void abundance to these methodological differences.

Due to the limited formation of biased tracers at early times, halo voids are consequently fewer in numbers. These early halos represent the highest peaks in the cosmic density field, and many halos that will later reside in clusters, filaments, and voids have not yet formed. As the redshift decreases, more halos form and cross the chosen mass cuts, revealing additional structures and splitting the few high-redshift voids, thereby increasing their total number.

In \mr{}, both halo and void numbers steadily increase as halos form and accrete mass. However, in \hr{}, halo numbers peak at $z = 0.34$ and then slightly decrease due to fewer new halos forming and some merging. Similarly, halo void numbers in \hr{} peak at a slightly higher redshift ($z = 0.67$) and decrease as some void boundaries merge, likely due to halos in filaments moving toward clusters. The trends depicted in Figure~\ref{fig_tracer_void_numbers} suggest that in the future, CDM and halo void numbers could converge, or at least become even closer than at present, as more CDM accretes in halos, potentially dissolving some current void boundaries.

\begin{figure}[!htbp]
    \centering
    \begin{tabular}{@{}c@{}}
        \includegraphics[width=0.82\linewidth, trim=0 7 0 4, clip]{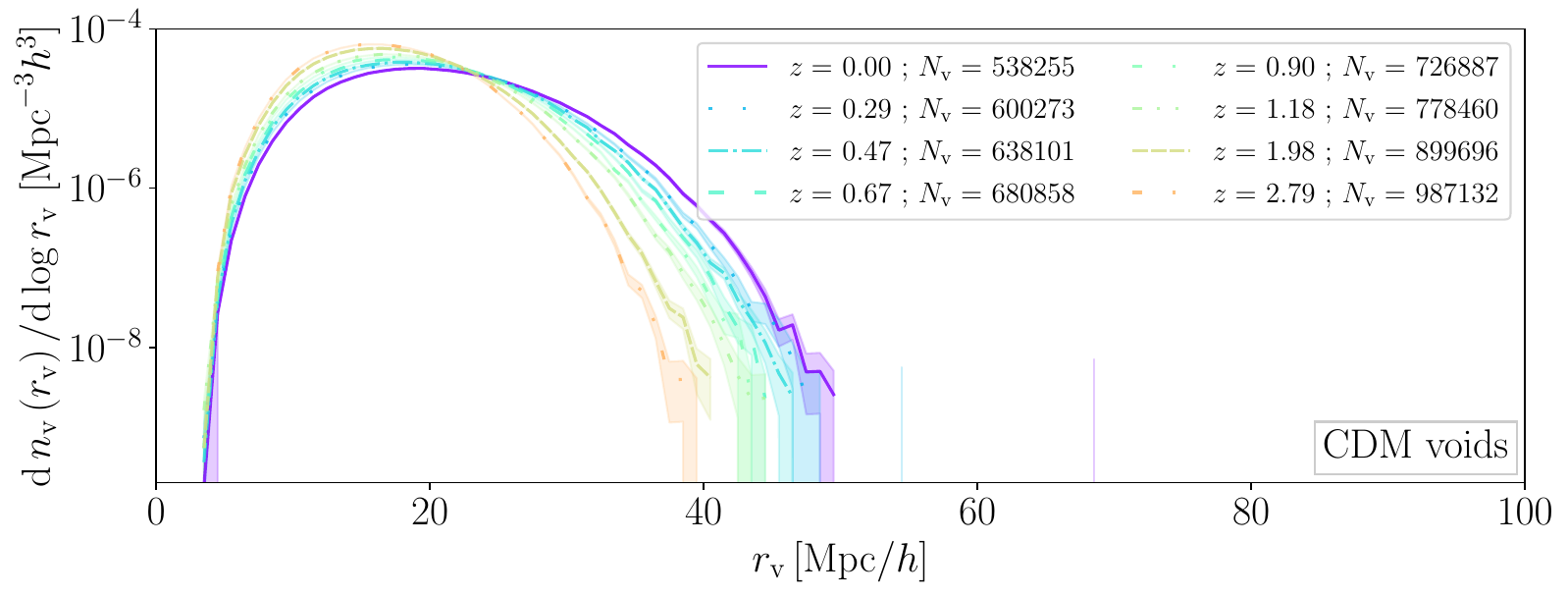} \\
        \includegraphics[width=0.82\linewidth, trim=0 7 0 4, clip]{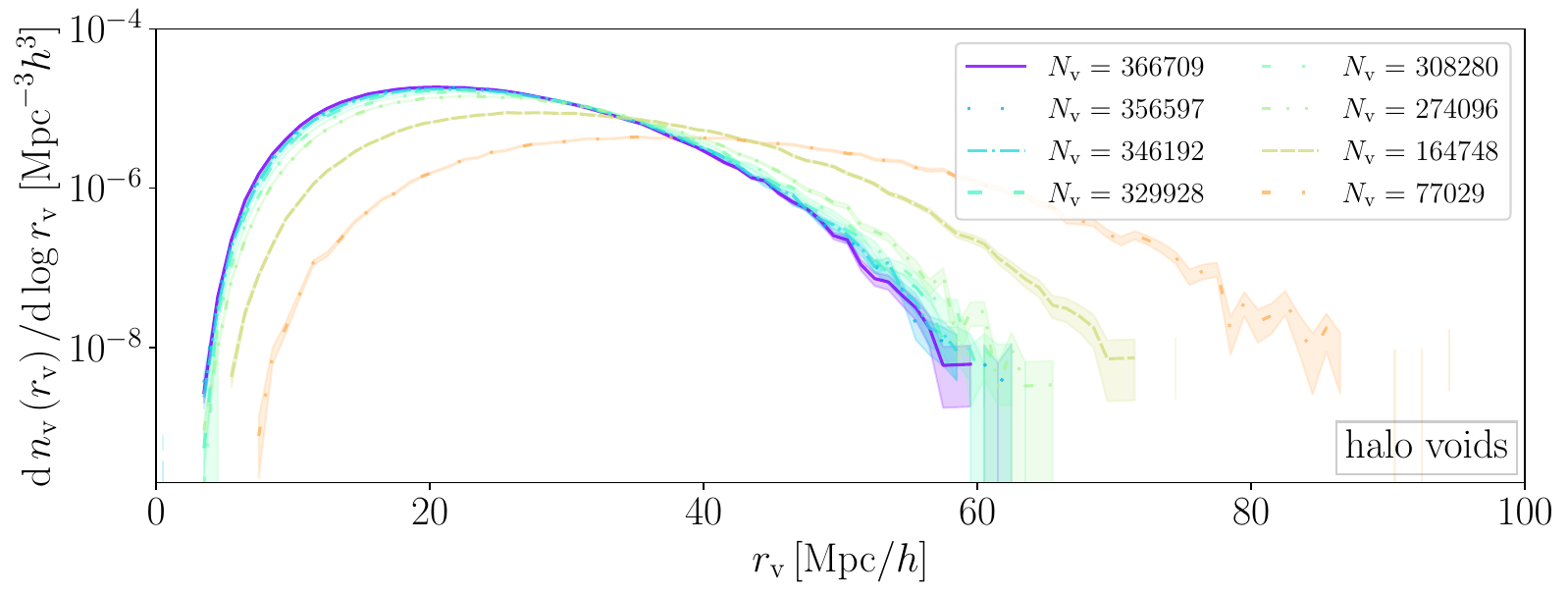} \\
        \begin{minipage}{0.9\linewidth}
            \small\raggedright
            {\textbf{(a)}} VSF evolution in \MR{} for CDM voids (top) and halo voids (bottom).
        \end{minipage} \\[0.4em]
        \includegraphics[width=0.82\linewidth, trim=0 7 0 4, clip]{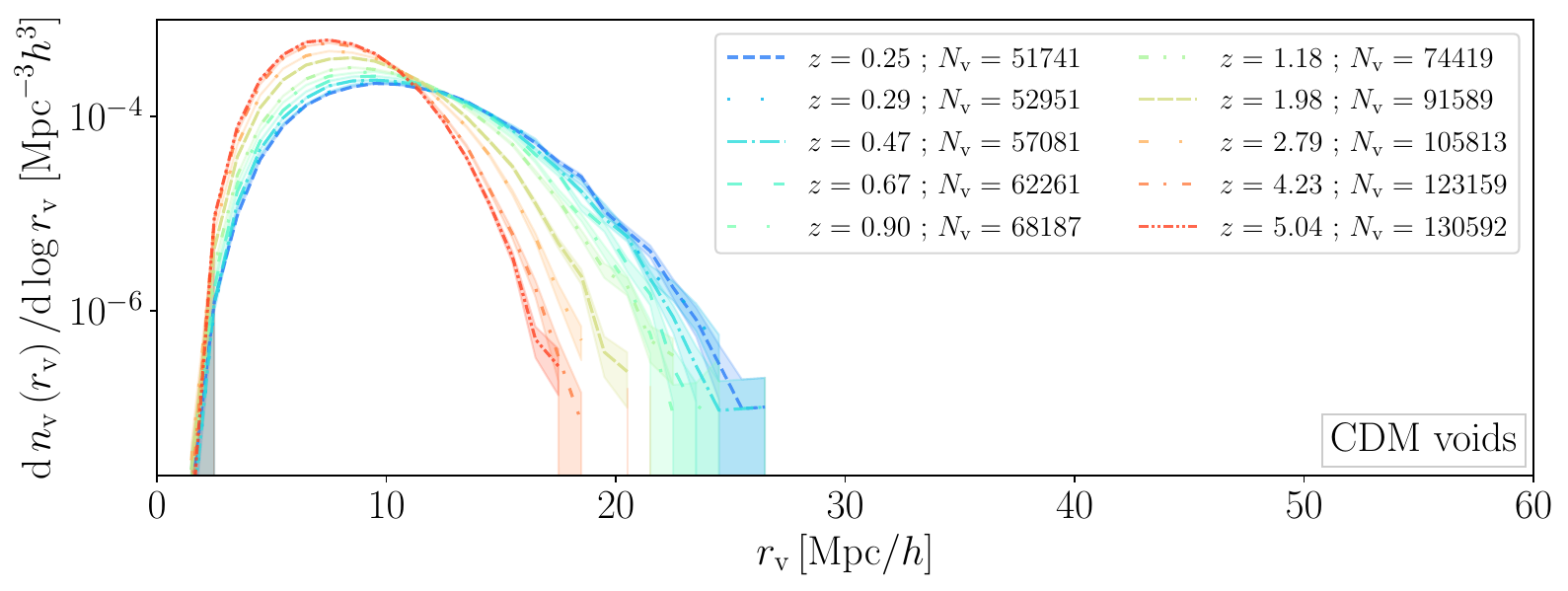} \\
        \includegraphics[width=0.82\linewidth, trim=0 7 0 4, clip]{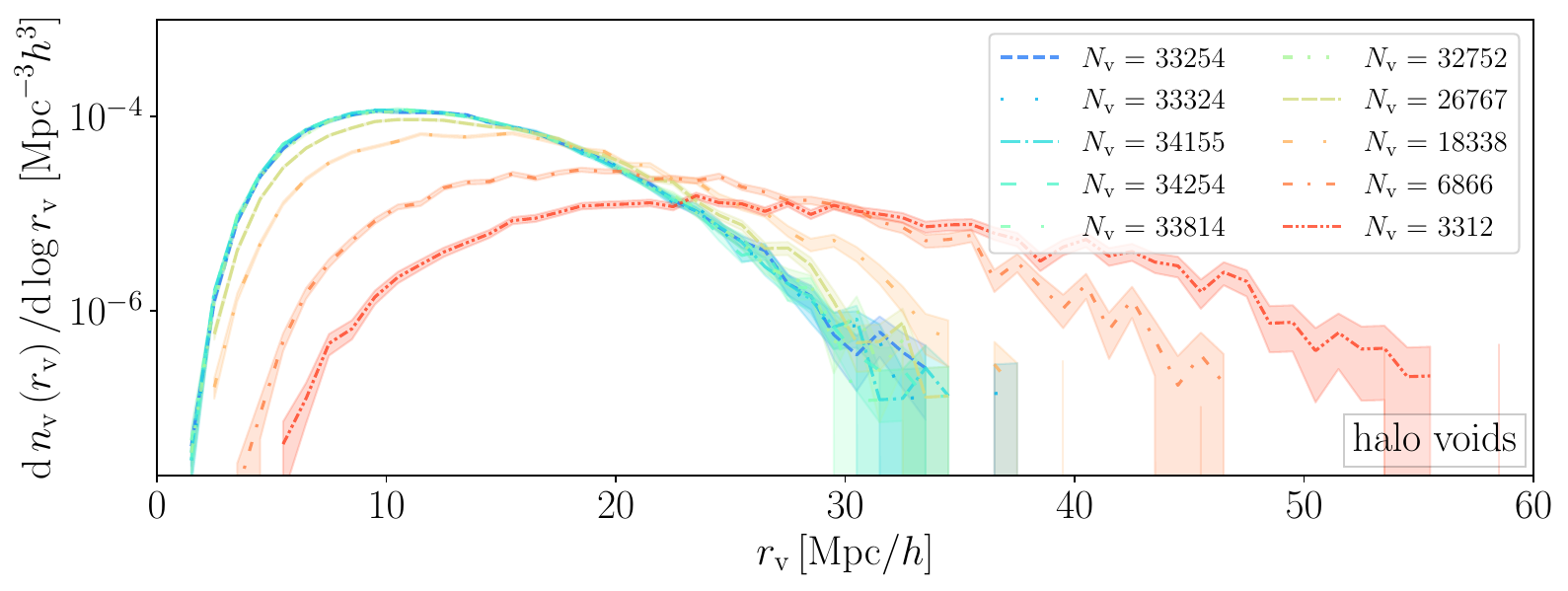} \\
         \begin{minipage}{0.9\linewidth}
            \small\raggedright
            {\textbf{(b)}} VSF evolution in \HR{} for CDM voids (top) and halo voids (bottom).
        \end{minipage}
    \end{tabular}
    \caption{Evolution of the void size function (VSF) for voids in \mr{}~\textbf{(a)} and \hr~\textbf{(b)}. Redshifts and respective void numbers at each snapshot are given in the legends. CDM voids grow larger over time through merging, while halo voids fragment into smaller voids before their comoving size distribution stabilizes at late times ($z \lesssim 1$). This demonstrates that late-time halo void evolution is no longer shaped by ongoing halo formation and mergers, but is instead primarily driven by cosmic expansion.}
    \label{fig_void_size_functions}
\end{figure}

\begin{figure}[!htbp]
    \centering
    \begin{tabular}{@{}c@{}}
        \begin{tabular}{@{}c@{}c@{}c@{}}
            \includegraphics[width=0.333\textwidth, trim=0 0 0 7, clip]{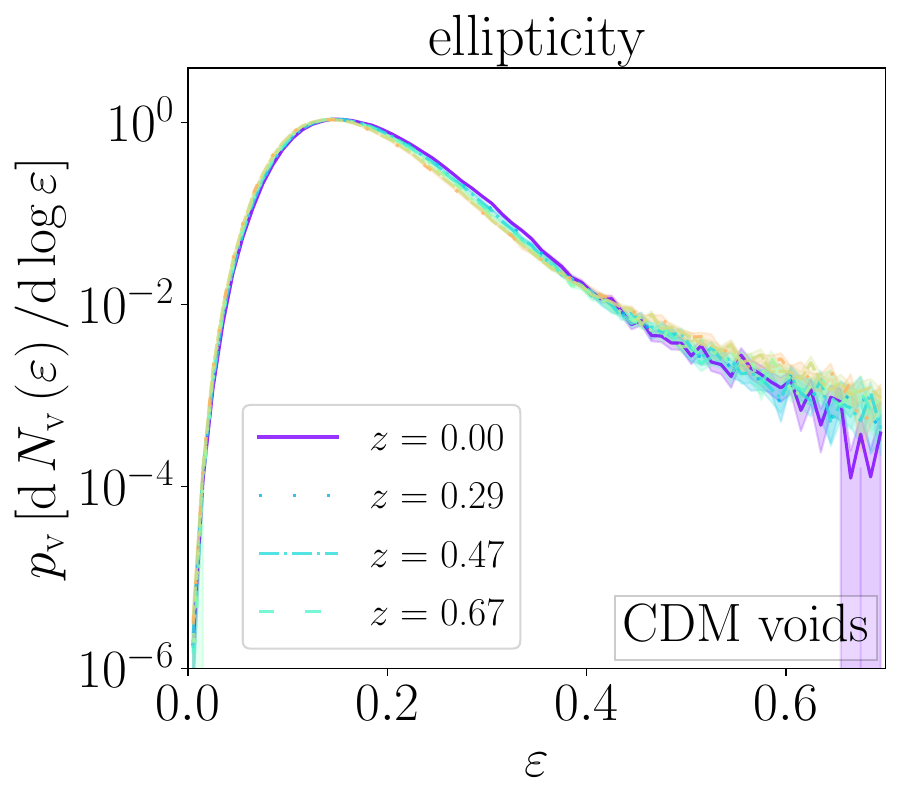}%
            \includegraphics[width=0.333\textwidth, trim=0 0 0 7, clip]{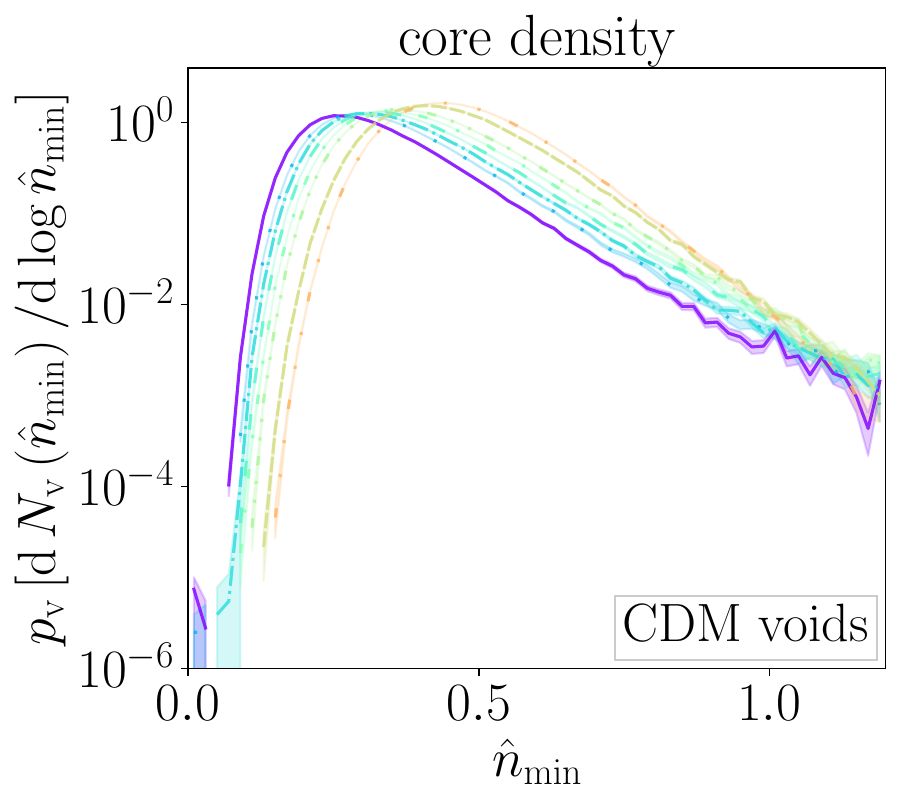}%
            \includegraphics[width=0.333\textwidth, trim=0 0 0 7, clip]{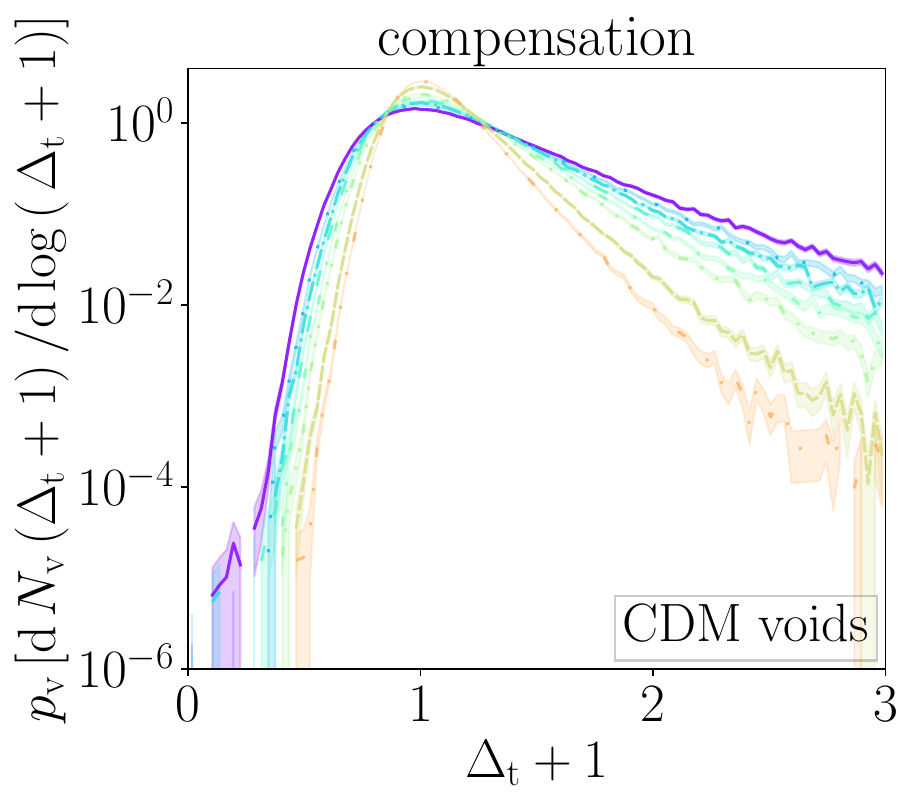} \\
            \includegraphics[width=0.333\textwidth]{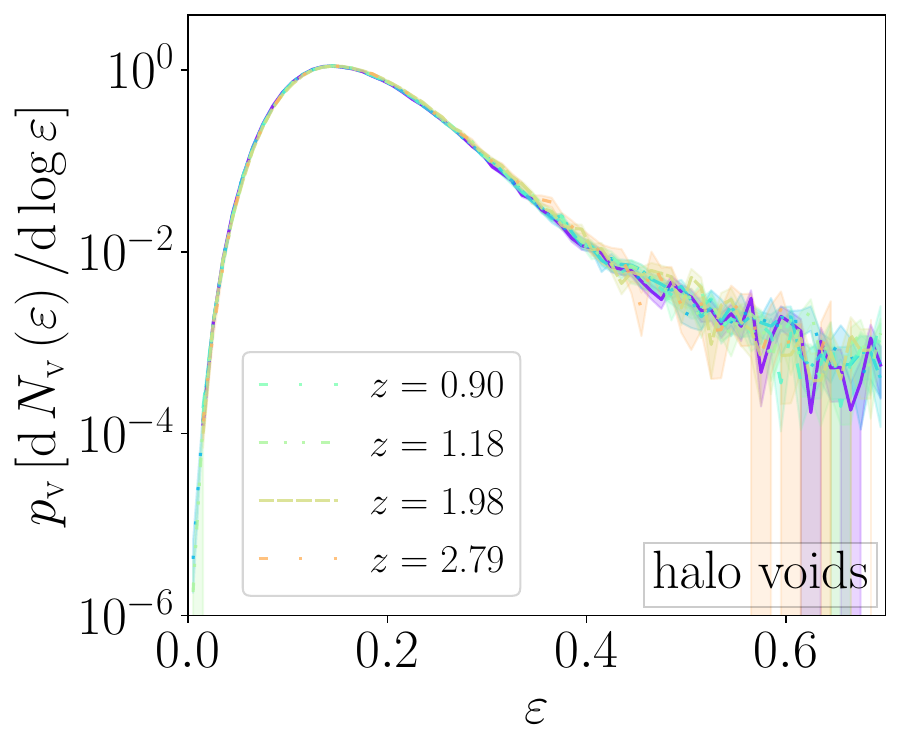}%
            \includegraphics[width=0.333\textwidth]{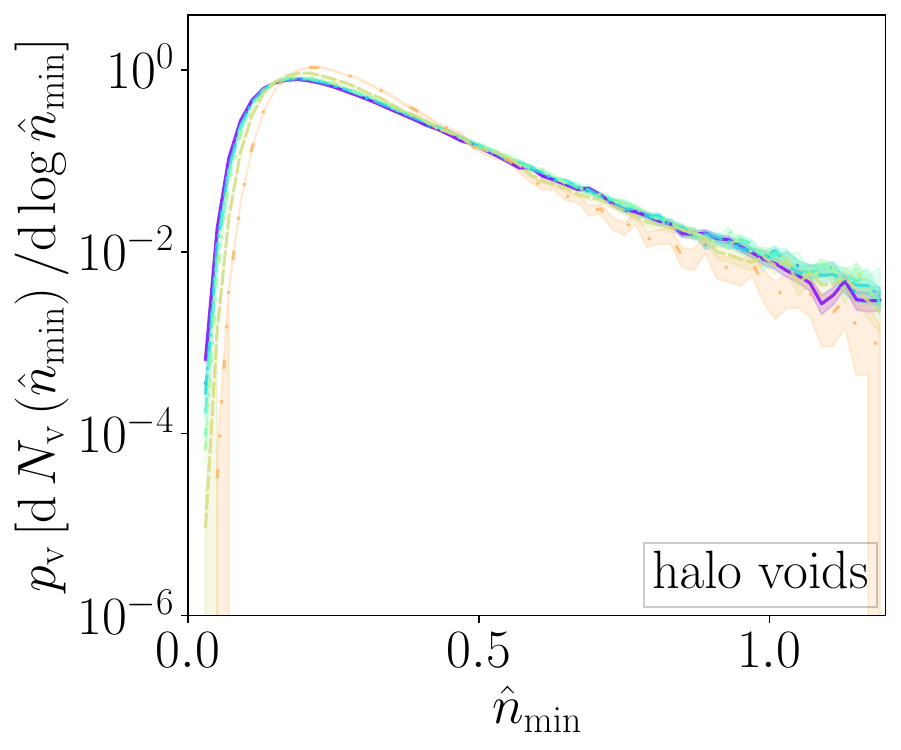}%
            \includegraphics[width=0.333\textwidth]{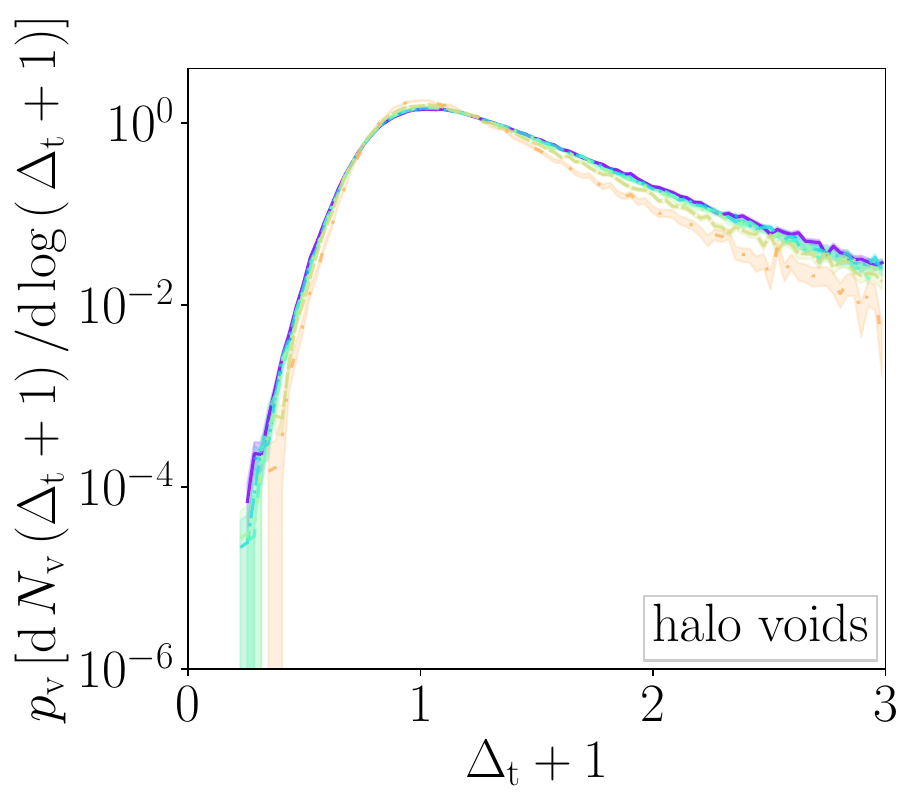}
        \end{tabular} \\
        \begin{minipage}{0.9\linewidth}
            \small\raggedright
            {\textbf{(a)}} Evolution of void properties in \MR{} for CDM voids (top) and halo voids (bottom).
        \end{minipage} \\[1em]
        \begin{tabular}{@{}c@{}c@{}c@{}}
            \includegraphics[width=0.333\textwidth]{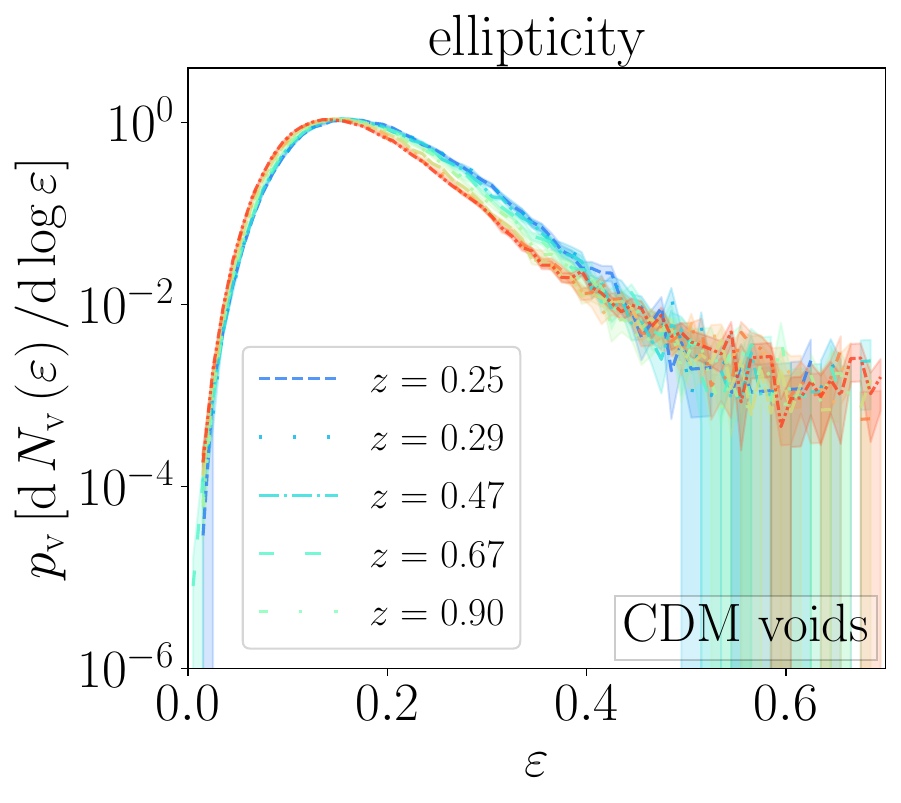}%
            \includegraphics[width=0.333\textwidth]{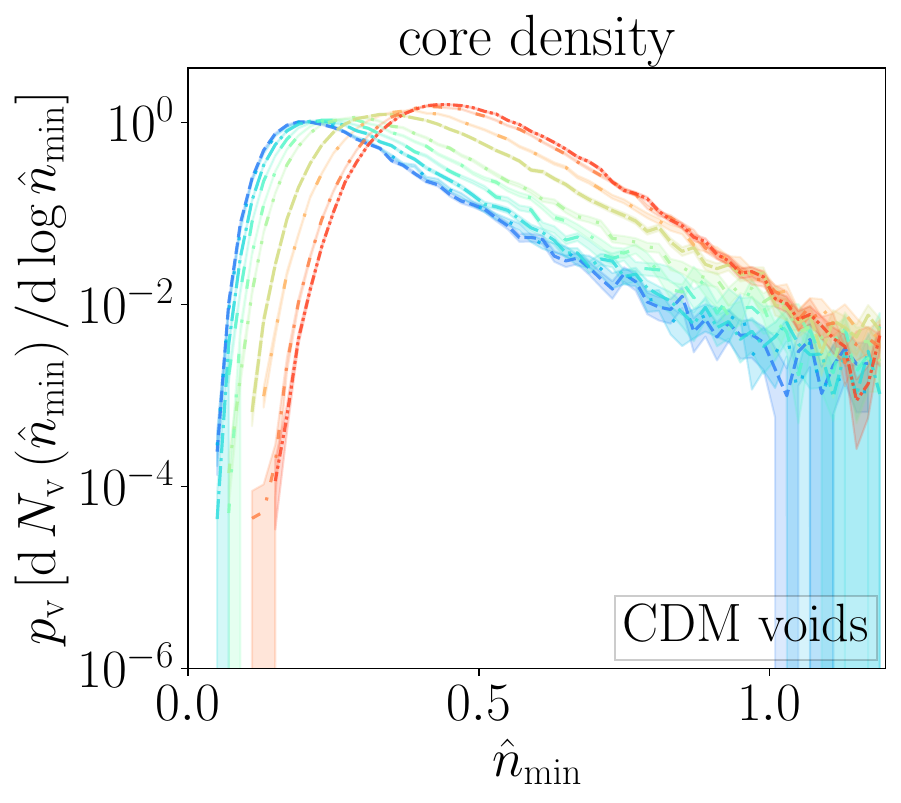}%
            \includegraphics[width=0.333\textwidth]{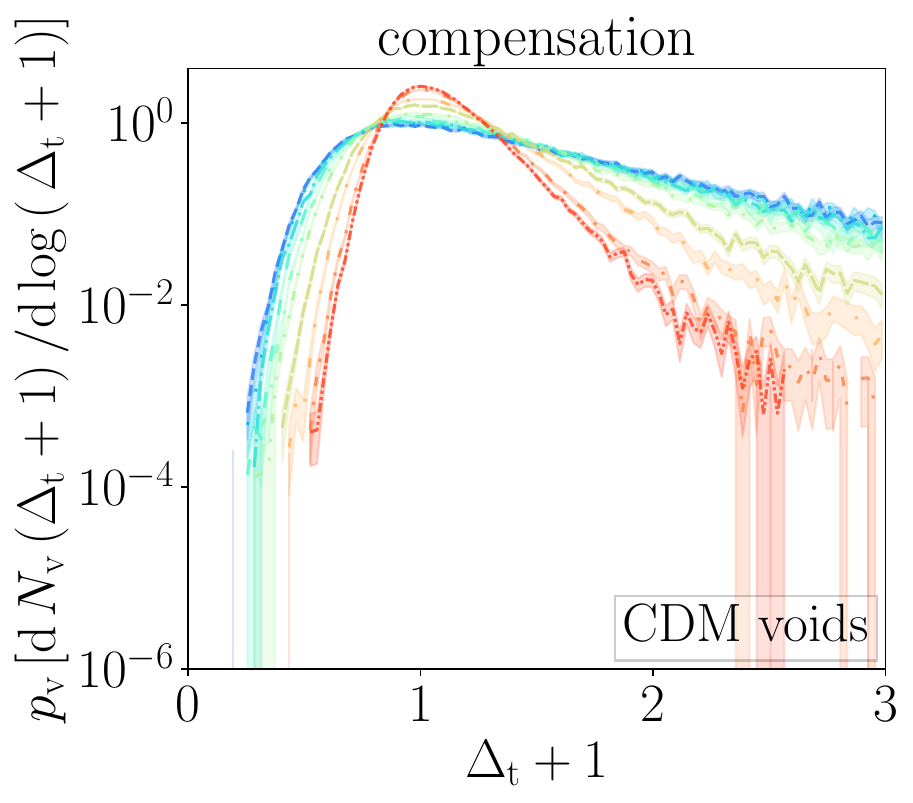} \\
            \includegraphics[width=0.333\textwidth]{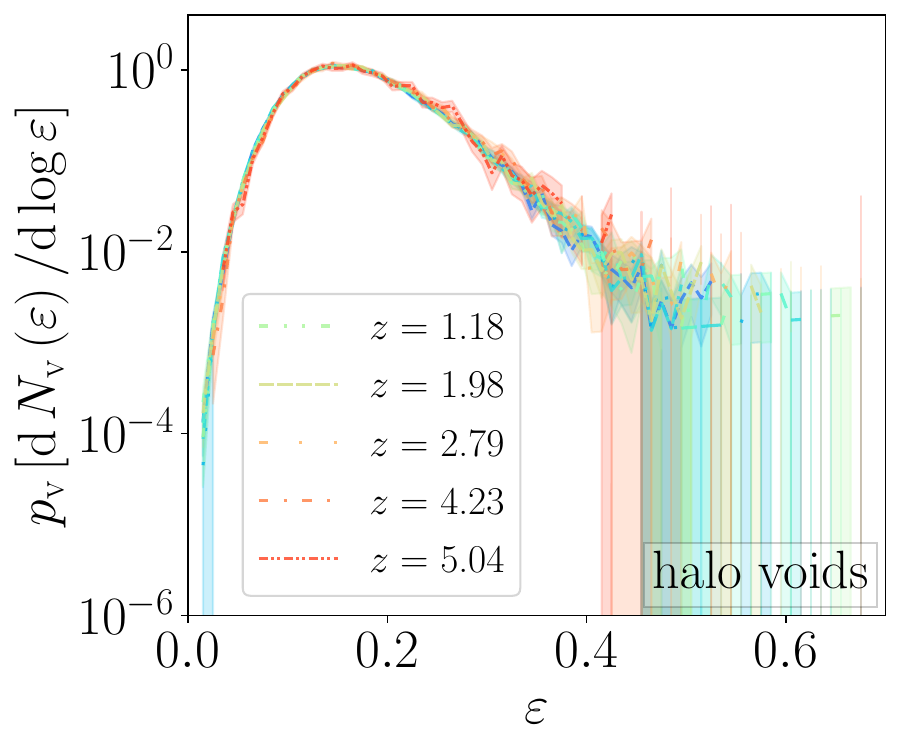}%
            \includegraphics[width=0.333\textwidth]{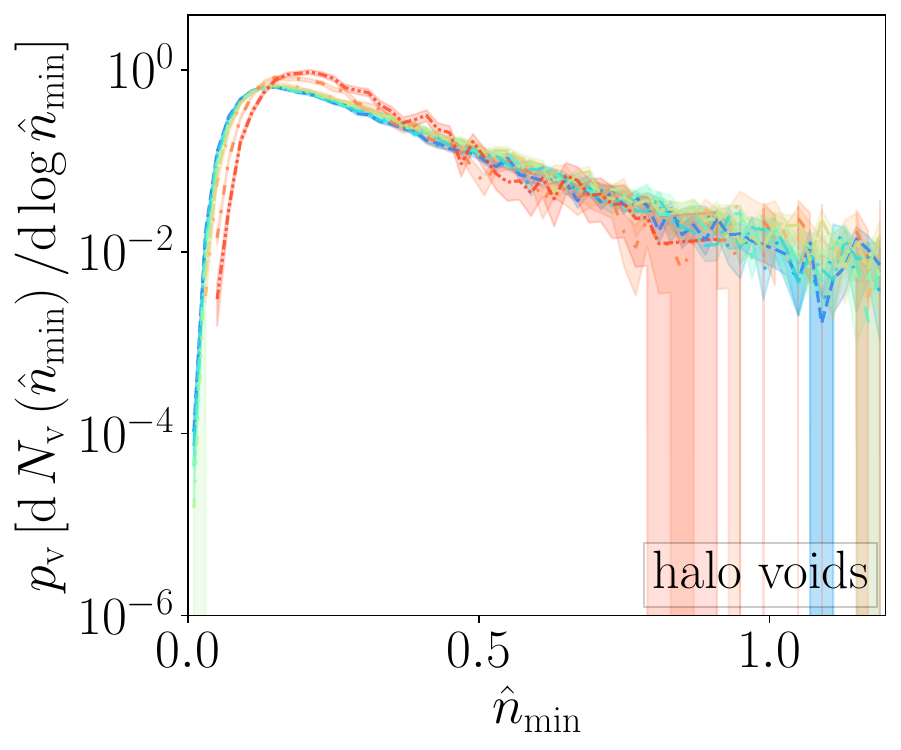}%
            \includegraphics[width=0.333\textwidth]{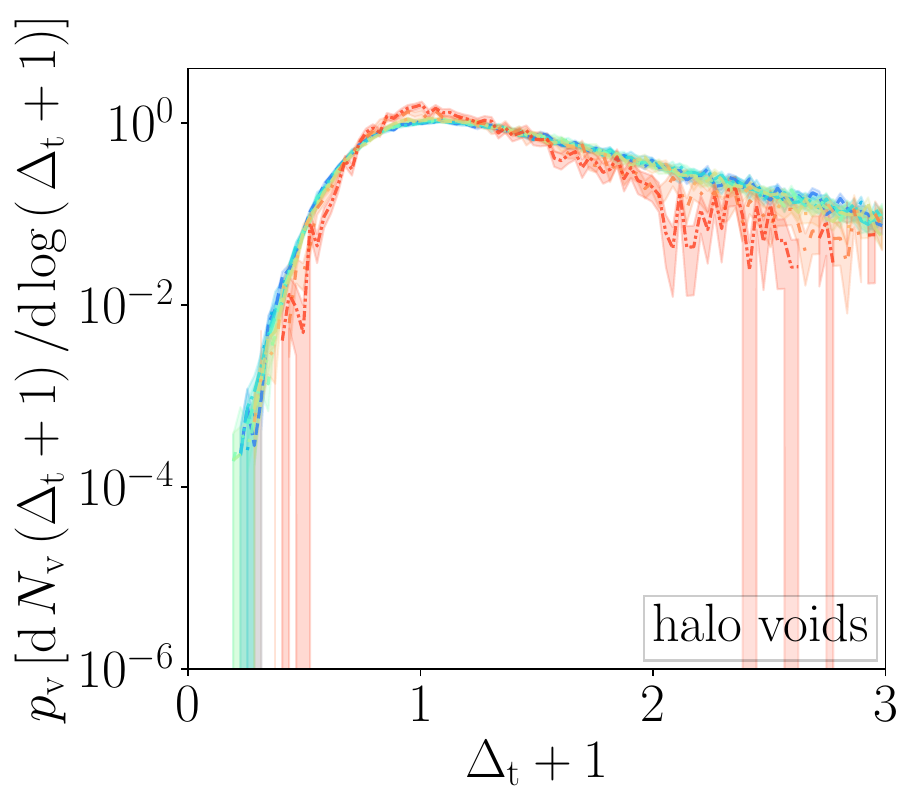}
        \end{tabular} \\
        \begin{minipage}{0.9\linewidth}
            \small\raggedright
            {\textbf{(b)}} Evolution of void properties in \HR{} for CDM voids (top) and halo voids (bottom).
        \end{minipage} \\[1em]
    \end{tabular}
    \caption{Evolution of void ellipticity (left column), core density (middle column) and compensation (right column) in \textbf{(a)}~\mr{} and \textbf{(b)}~\hr{}. Redshifts are indicated in the legends of the left column. All distributions are normalized by the total void count at each redshift to account for the evolving number of voids.}
    \label{fig_void_property_evolution}
\end{figure}

\subsection{The evolution of void properties \label{subsec:void_properties}}

Figure~\ref{fig_void_size_functions} presents the evolution of the void size function (VSF), while Figure~\ref{fig_void_property_evolution} shows the evolution of other common void properties: ellipticity (left), core density (middle), and compensation (right). Both figures compare CDM and halo voids in the \MR{} simulation (first two rows) and in the \HR{} simulation (last two rows).

For the distribution of void radii in Figure~\ref{fig_void_size_functions}, we observe distinct evolutionary paths for CDM and halo voids in both simulations. The differences are most pronounced at high redshifts, where numerous small CDM voids populate the dense tracer field, in stark contrast to the few, large halo voids identified in the sparse halo population. As the Universe evolves, their paths diverge: CDM voids grow larger and fewer in number through hierarchical merging, a process known as the `void-in-void' evolution~\cite{Sheth2004}. Meanwhile, the initially large halo voids are progressively split into smaller structures as new halos form and populate the cosmic web, strengthening earlier results~\cite{Verza2022,Curtis2025}. For context, the mean halo separations at $z_\mathrm{min}$ are on the order of $6.7 \, \Mpch$ for \mr{} and $3.2 \, \Mpch $ for \hr{}.

A key new feature evident in both simulations is the stabilization of the halo void population at late times. The comoving VSFs in \hr{} align almost perfectly at redshifts $z \lesssim  1.32$, a pattern that appears later and is less pronounced in \mr{} at $z \lesssim 0.67$. This alignment indicates that the evolution of halo voids is no longer shaped by the ongoing formation and merging of halos, but is instead driven primarily by cosmic expansion.

Given that void size evolution is directly linked to the number of identified voids in a fixed comoving volume, we aim to isolate the evolution of other void properties independently of these number changes. To this end, Figure~\ref{fig_void_property_evolution} displays the void property evolution, with each distribution normalized to the void count at a given redshift.

Unlike their sizes, void shapes, represented by their ellipticities (left column), evolve less significantly over time, supporting prior findings~\cite{Wojtak2016,VallesPerez2021}. For CDM voids, presented in the upper row for each simulation, the median ellipticities and the maximum in the distribution slightly increase at lower redshifts, a trend that is more prominent in the \hr{} simulation due to the smaller scales being probed. This seemingly contradicts common models on isolated void evolution~\cite{Sheth2004}, but the merging of CDM voids over time offers a natural explanation, as it allows for the formation of more complex shapes with higher ellipticities. In contrast, the distributions for halo voids, shown in the lower rows, remain largely constant, demonstrating only a minuscule hint of more elliptical voids at high redshifts. This suggests that the initially irregular shapes of early halo voids, likely due to a random distribution of the first density peaks, simplify into more spherical structures as they fragment over time. Voids with ellipticities  $\varepsilon > 0.7$ are excluded from Figure~\ref{fig_void_property_evolution} due to their extremely low numbers.

The core density distributions in the middle column of Figure~\ref{fig_void_property_evolution} illustrate how CDM voids (upper row of each pair) become increasingly devoid of matter over time~\cite{Sutter2014c,Wojtak2016,VallesPerez2021}. Early CDM voids possess high core densities that substantially decrease as matter accretes and migrates outwards towards their boundaries. By comparison, the core densities of halo voids evolve far less, with their median values remaining nearly stationary over a wide range of cosmic time. Small shifts towards higher densities are observed only at high redshifts ($z \geq 2.79$), becoming slightly more prominent in \hr{} at $z \geq 4.23$. Even so, these shifts are still significantly smaller than for CDM voids, and the absolute values of $\coreDens$ are resolution-dependent. Notably, the core density distributions for halo voids begin to align at approximately the same redshifts as their VSFs, indicating a linked stabilization.

This divergence in core density evolution between CDM and halo voids is physically expected. CDM void centers become increasingly underdense as matter moves outwards without new particle creation. Conversely, halo voids exhibit minimal core density changes because, even at early times, only a few (and typically less massive) halos reside near their centers. These central halos likely formed later and grow more slowly due to limited accretion compared to halos near void boundaries. As halo voids evolve, existing halos continue moving outwards, but this outward motion is largely balanced by new halos forming and crossing the mass threshold inside voids, resulting in only minor changes to core density values.

Lastly, we explore the evolution of void compensations (right column of Figure~\ref{fig_void_property_evolution}). The medians of the distributions for both CDM and halo voids remain remarkably constant over time. This supports the view that a void's compensation is established early and rarely transitions between under- and overcompensation. This is expected, as the compensation is largely predetermined by the initial cosmic density field~\cite{Hamaus2014a,Hamaus2014b}. This stability is maintained even as the void populations evolve differently: as halo voids split, the fragments inherit the compensation of their predecessor, while merging CDM voids combine the similar compensation values of their environmentally-linked predecessors.

While the medians are stable, the overall distributions do evolve~\cite{Adermann2018}. The mean compensations for both void types trends towards slightly higher values due to an asymmetric evolution: matter accretion from void interiors and regions near their boundary can indefinitely amplify the density of already overdense regions, while the evolution of underdense regions is limited by the physical lower bound of $\Delta_\tracer = -1$ ($\rho = 0$). This effect, which drives the mean upwards, is most visible for CDM voids, whose compensation distribution broadens considerably over time, a trend that is attenuated for halo voids.

The evolutionary trends of void properties presented in this section reveal a remarkable self-similarity across scales between the \MR{} and \HR{} simulations. The different resolutions and mass cuts effectively probe the same physical evolution at different scales and cosmic epochs: the evolutionary stages of small voids at high redshift in \hr{} are mirrored by those of larger voids at lower redshift in \mr{}. This is clearly visible in the property distributions (Figures~\ref{fig_void_size_functions} and~\ref{fig_void_property_evolution}), which depict nearly identical trends, simply shifted in scale and time. Additionally, we confirm these evolutionary trends for CDM voids in the \hr{} simulation when using subsamples of higher tracer density to probe their evolution on even smaller scales. This finding, which confirms the self-similar nature of voids seen in previous work~\cite{Schuster2023,Schuster2024}, motivates a deeper look into their internal structure. We therefore now turn to the evolution of their density profiles to see if this scale-invariance holds.

\section{Evolution of density profiles \label{sec:density_profiles}}

To see if the self-similarity observed in global void properties extends to their internal structure, we now investigate the evolution of their density profiles. Beyond testing for this universality, the evolution of these profiles serves as a fundamental test of cosmological models and our understanding of how structures change in the Universe's most underdense regions.

We will compare the profiles of voids identified in CDM tracers with those identified in halos. For the latter, we also analyze the distribution of both CDM and the halos themselves. Since individual halo voids cannot be reliably tracked over time, our analysis focuses on profiles stacked in bins of void radius $r_\void$, always identifying voids and calculating their profiles independently at each redshift. Unless stated otherwise, these profiles are calculated as described in Section~\ref{sec:methods} in spherical shells of width $0.1 \times r_\void$ and depicted in units of $r / r_\void $ for easier comparisons. To ensure robust and more detailed density estimations in the \HR{} simulation, CDM densities are calculated using a higher-density CDM subsample ($5 \times 10^7$ particles) to reduce noise from sparse sampling in void interiors and to probe their true substructure more accurately.

\subsection{CDM and halo voids: a tale of two tracers \label{subsec:CDM_halo_voids}}

\begin{figure}[t!]
               \centering \resizebox{\hsize}{!}{
                               \includegraphics[trim=0 0 0 0, clip]{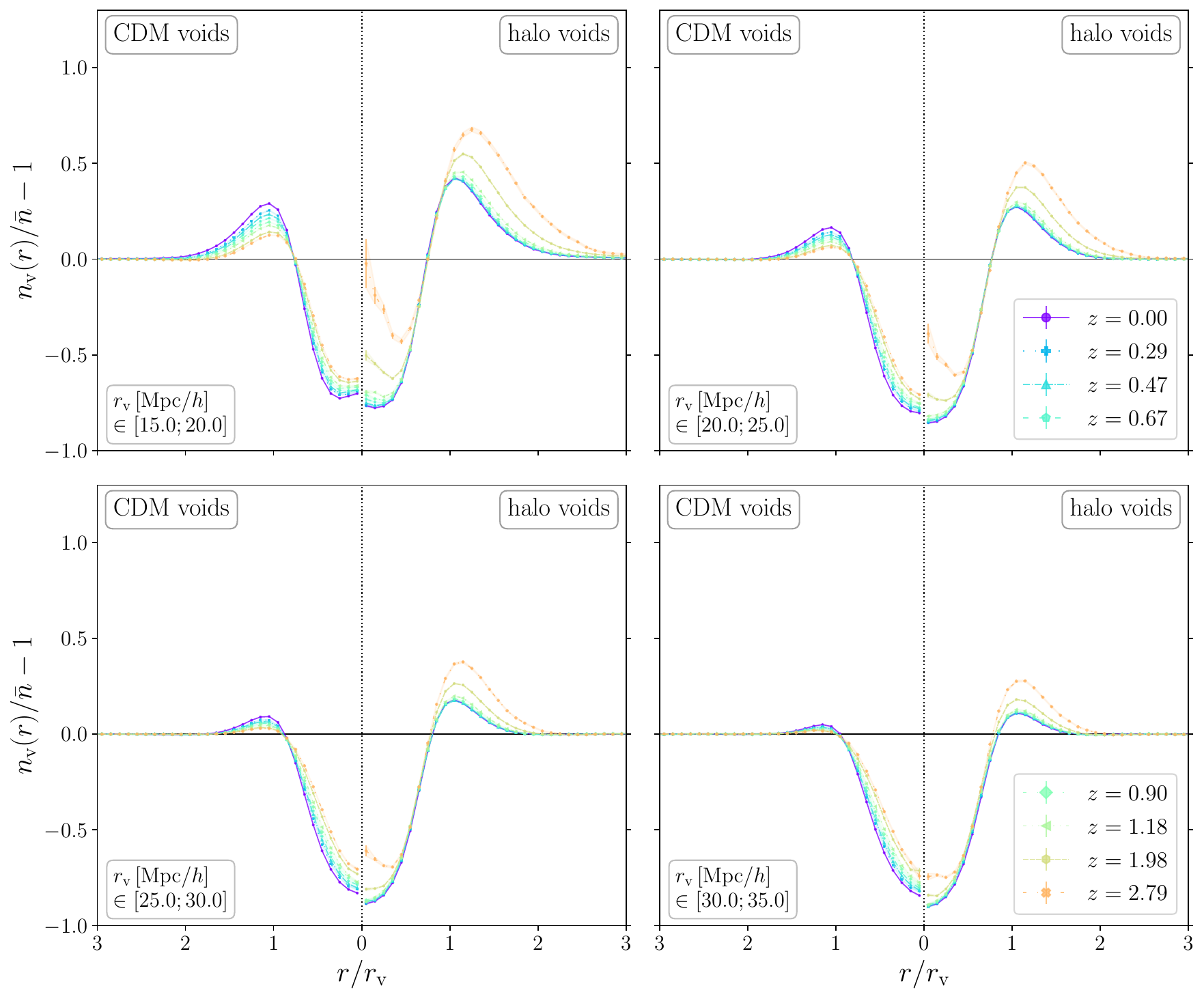}}
               \caption{Density profiles of \emph{isolated} CDM voids (left side of each panel) and halo voids (right side of each panel) from the \MR{} simulation at different redshifts, indicated in the legends on the right side of each row. All profiles are stacked in bins of void radius, with the bin edges given on the bottom left of each panel. While CDM voids evolve as expected, halo voids show an apparent `inverse' evolution in fixed-size bins, a selection effect we address later.}
               \label{fig_density_voids_midres_radius}
\end{figure}

\begin{figure}[t!]
               \centering \resizebox{\hsize}{!}{
                               \includegraphics[trim=0 0 0 0, clip]{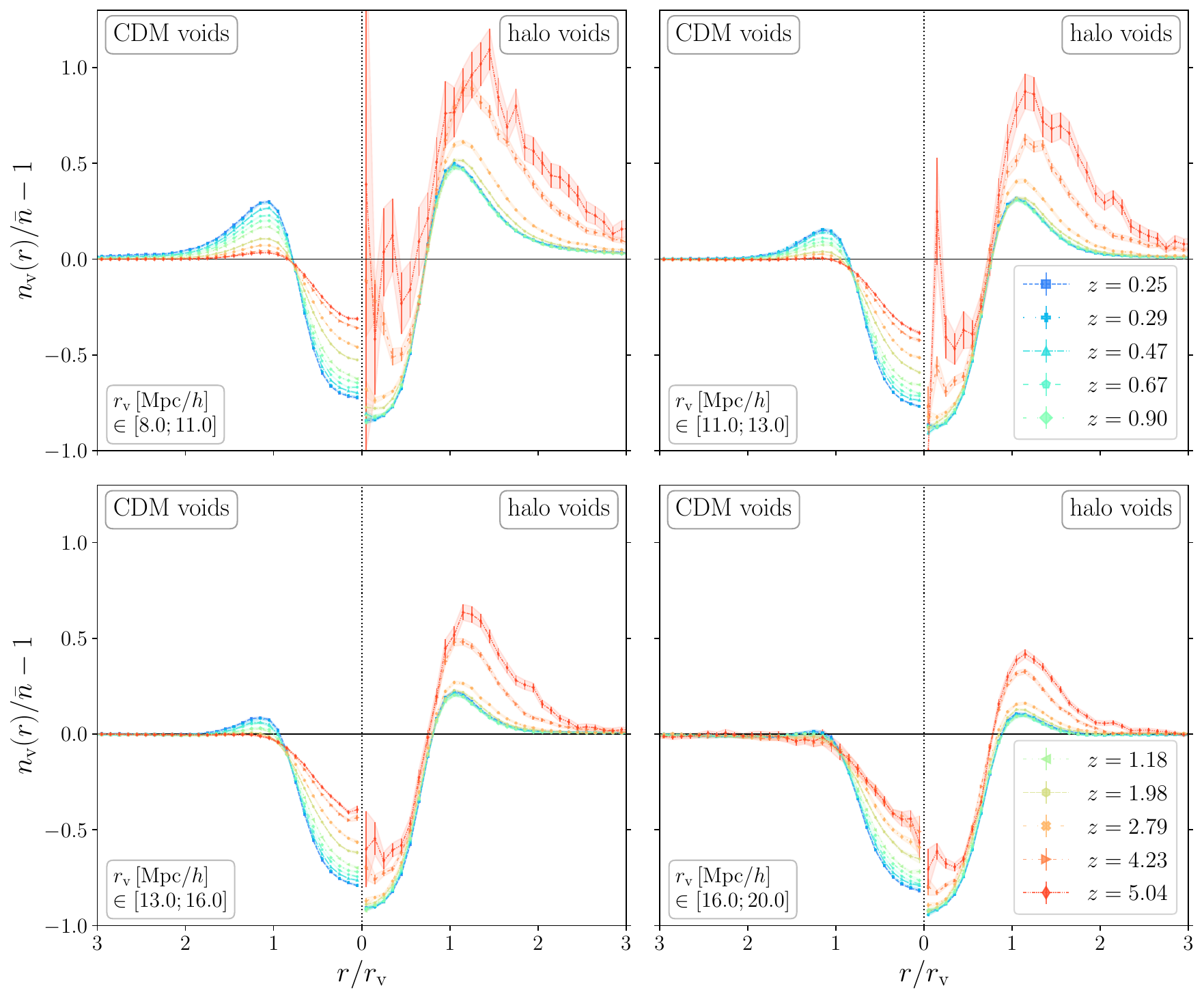}}
               \caption{Same as Figure~\ref{fig_density_voids_midres_radius}, but for the \HR{} simulation. To better resolve the inner structure of CDM voids, their density profiles are calculated using a higher-density tracer subsample.}
               \label{fig_density_voids_highres_radius}
\end{figure}

\begin{figure}[t!]
               \centering \resizebox{\hsize}{!}{
                               \includegraphics[trim=0 0 0 0, clip]{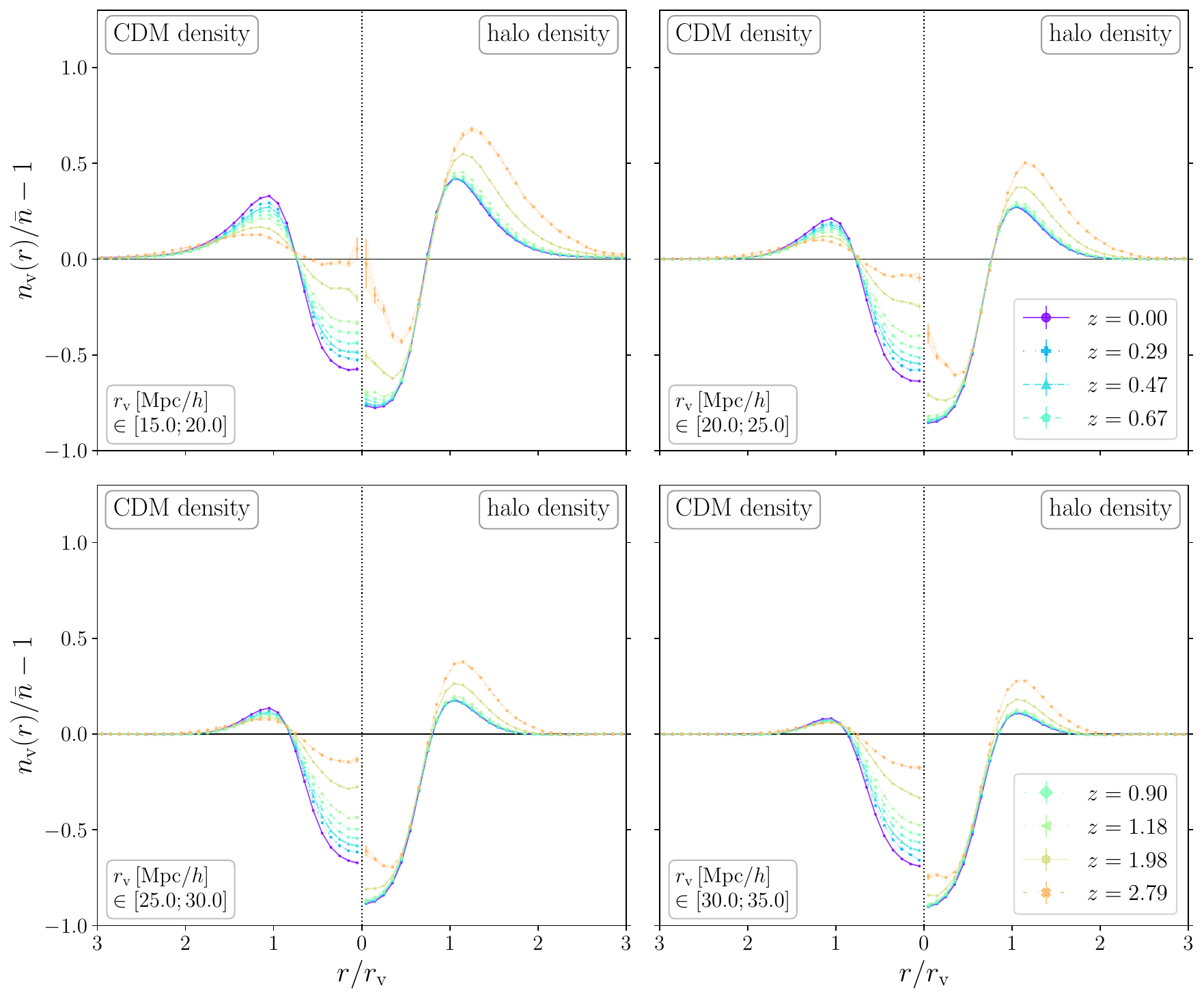}}
               \caption{Density profiles of CDM and halos around halo voids from the \MR{} simulation, shown on the left and right sides of each panel, respectively. Redshifts are indicated in the legends on the right side of each row. All profiles are stacked in bins of void radius, with the bin edges given on the bottom left of each panel. This comparison reveals that the apparent inverse evolution of halo voids is a selection effect, since the underlying CDM density evolves as physically expected.}
               \label{fig_density_halo_voids_CDM_midres_radius}
\end{figure}

\begin{figure}[t!]
               \centering \resizebox{\hsize}{!}{
                               \includegraphics[trim=0 0 0 0, clip]{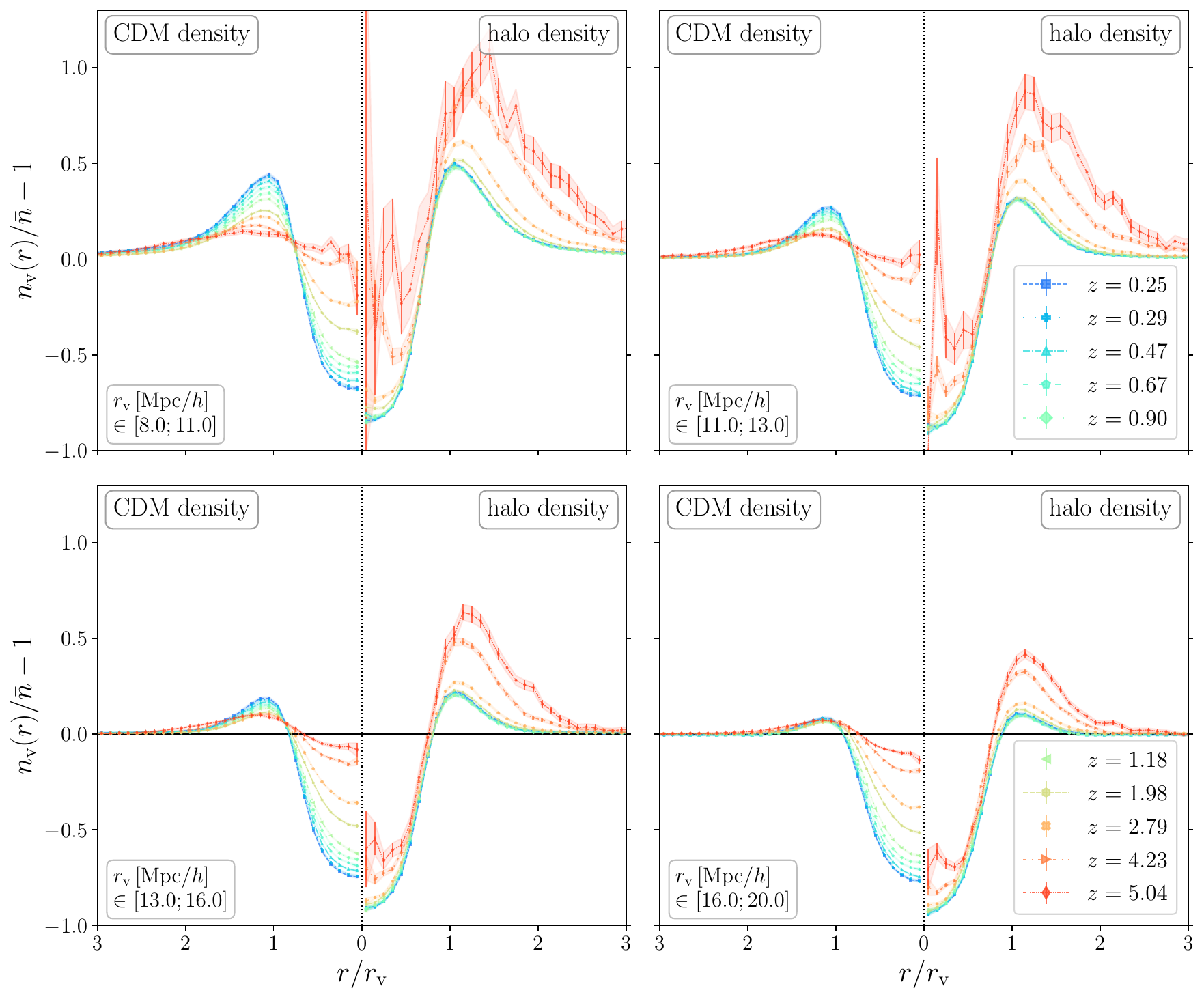}}
               \caption{Same as figure~\ref{fig_density_halo_voids_CDM_midres_radius}, but for the \HR{} simulation, using a denser CDM subsampling for improved accuracy in the void interiors}
               \label{fig_density_halo_voids_CDM_highres_radius}
\end{figure}

Figures~\ref{fig_density_voids_midres_radius} and~\ref{fig_density_voids_highres_radius} compare the number density profiles of voids identified in CDM (left panels) and halos (right panels) in the \MR{} and \HR{} simulations, respectively. To ensure a direct comparison, we use identical void radius bins for both tracer types. The CDM voids in both simulations follow the expected trend: as matter streams from their interiors to their boundaries over time, they become emptier and their compensation walls grow, consistent with previous work~\cite{Sheth2004, Cautun2014, Hamaus2014b, Sutter2014c, Wojtak2016, Massara2018} and the evolutionary trends in core density discussed in Section~\ref{subsec:void_properties}.

A notable difference is that the profiles of CDM voids in \hr{} appear to have denser interiors, which seemingly contradicts both the core density measurements and the broader finding of self-similarity between simulations. However, follow-up tests show this is a methodological effect. We find that voids identified in a denser CDM subsample are themselves emptier, and that the apparent density differences vanish when the original \hr{} profiles are recalculated using only the CDM tracers from the void identification. The discrepancy is caused by using a denser tracer sample for the profile calculation than for the void identification, which raises the spherically-averaged inner density by better resolving internal substructure and smoothing out subvoids present in denser samples. Accounting for this measurement effect restores the expected self-similarity between the simulations, while leaving the compensation walls and exterior profiles unaffected.

With the behavior of CDM voids established, we now turn to the more complex evolution of voids identified in the biased halo population in the right panels of Figures~\ref{fig_density_voids_midres_radius} and~\ref{fig_density_voids_highres_radius}. While void interiors still follow the expected evolution of voids becoming emptier, the innermost regions at high redshift can show signs of sparse sampling where densities are overestimated due to tracers occupying small-volume shells near void centers (see also~\cite{Schuster2023}). In contrast, their exteriors reveal an apparent `inverse' evolution: the compensation walls are most pronounced at high redshift and seem to decrease over time. To understand this, Figures~\ref{fig_density_halo_voids_CDM_midres_radius} and~\ref{fig_density_halo_voids_CDM_highres_radius} compare the halo number density (right panels) with the density of the underlying matter traced by CDM (left panels) around the same halo voids.

A clear contrast between the tracers is evident. At high redshift, the sparsely populated halo samples carve out voids with extremely deep interiors and highly pronounced compensation walls. In comparison, the underlying CDM density field is much smoother, showing only minor underdensities with interiors sometimes approaching the mean density and with less prominent compensation walls. The crucial difference lies in their evolution: while the halo profiles show the inverse trend, the underlying CDM density evolves as expected, with void interiors becoming progressively emptier as compensation walls grow through matter accretion~\cite{Curtis2025}. Notably, and as opposed to CDM void profiles, using a denser CDM subsample for the profile calculation does not affect the measured CDM densities around these halo-identified voids.

These findings confirm that the counterintuitive inverse evolution is a tracer-related selection effect, not a physical reversal of structure growth. Although the physical sizes of voids naturally increase with cosmic expansion, the binning in fixed comoving radii used here isolates their dynamic evolution. Because the halo void size function evolves significantly, a fixed-radius bin probes different void populations at different epochs, selecting the smallest, most highly compensated voids at high redshift, while selecting more average-sized voids at low redshift. This mixing of void populations creates the apparent inverse trend seen in the halo tracers, even while the core densities suggest the voids are becoming emptier over time. The halo bias further amplifies this effect, as the first halos preferentially form in the high-density peaks that define the void walls, whereas the underlying matter field is not subject to this strong bias.

With the selection effect understood, we can focus on the true physical evolution that becomes apparent at late times as the halo void population stabilizes. The comoving density profiles begin to align and their evolution slows significantly at the same redshifts where their size functions also show minimal change ($z \lesssim 0.67$ in \mr{} and $z \lesssim 1.32$ in \hr{}). This alignment signals a fundamental shift in the large-scale evolution of halo voids: the expansion of the Universe becomes the primary driver and their development over time is no longer dominated by the dynamic formation of new halos or the outflow of existing ones. This large-scale stability is complemented by a shift in the internal dynamics of the voids, where a dynamic equilibrium is reached. While voids are typically characterized by a net outflow of halos --- a trend we confirm in their evolving velocity profiles --- the stable inner halo densities at late times suggest this outflow is balanced by the formation of new halos within voids.

\subsection{A relative framework for void evolution \label{subsec:profiles_relative_bins}}

\begin{figure}[t!]
               \centering \resizebox{\hsize}{!}{
                               \includegraphics[trim=0 0 0 0, clip]{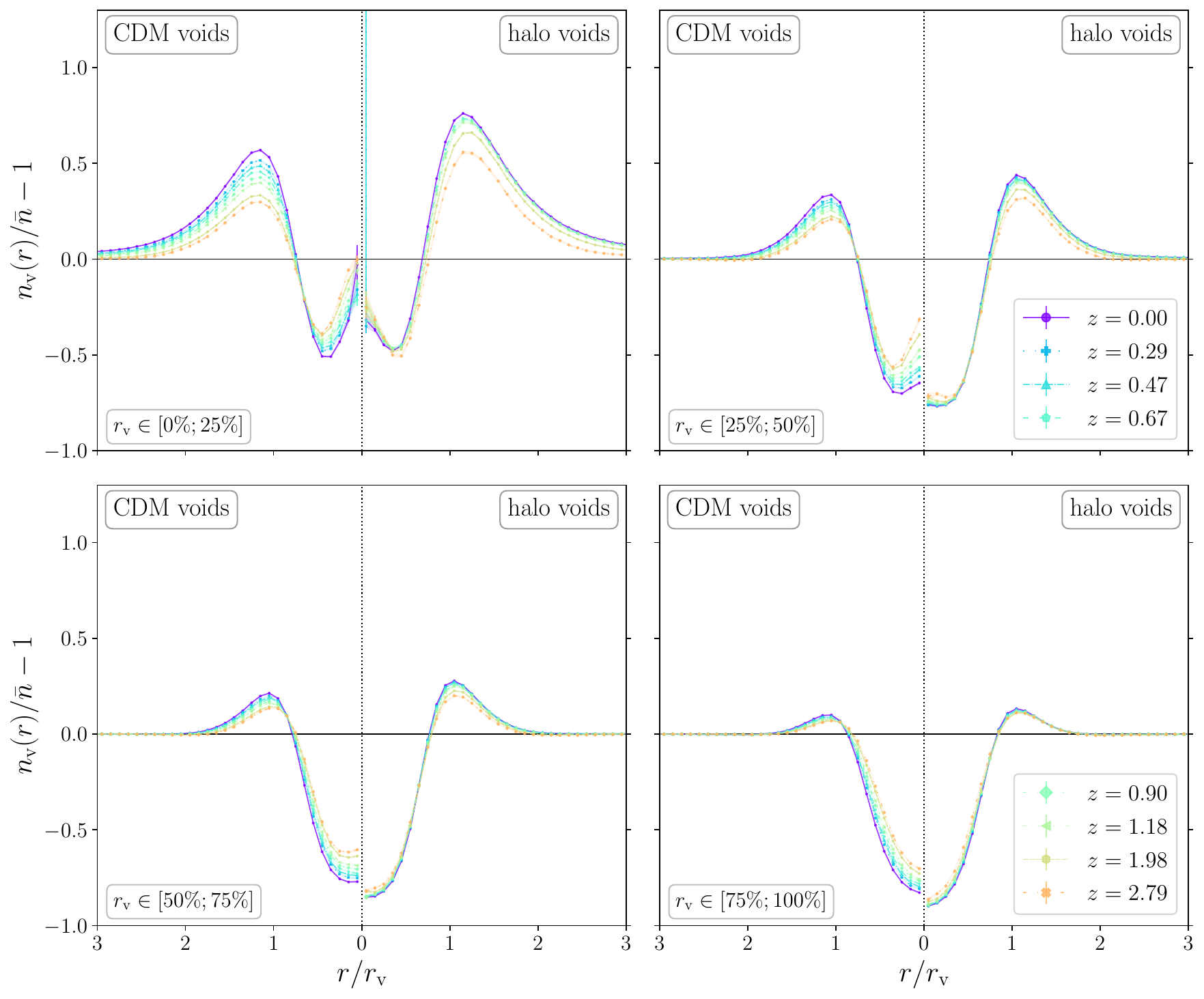}}
               \caption{Same as figure~\ref{fig_density_voids_midres_radius}, but with profiles stacked in percentile bins of void radius. By switching to this relative size framework, these profiles now reveal the true physical evolution of halo voids, with compensation walls growing over time, effectively removing the selection effect.}
               \label{fig_density_voids_midres_relative_radius}
\end{figure}

\begin{figure}[t!]
               \centering \resizebox{\hsize}{!}{
                               \includegraphics[trim=0 0 0 0, clip]{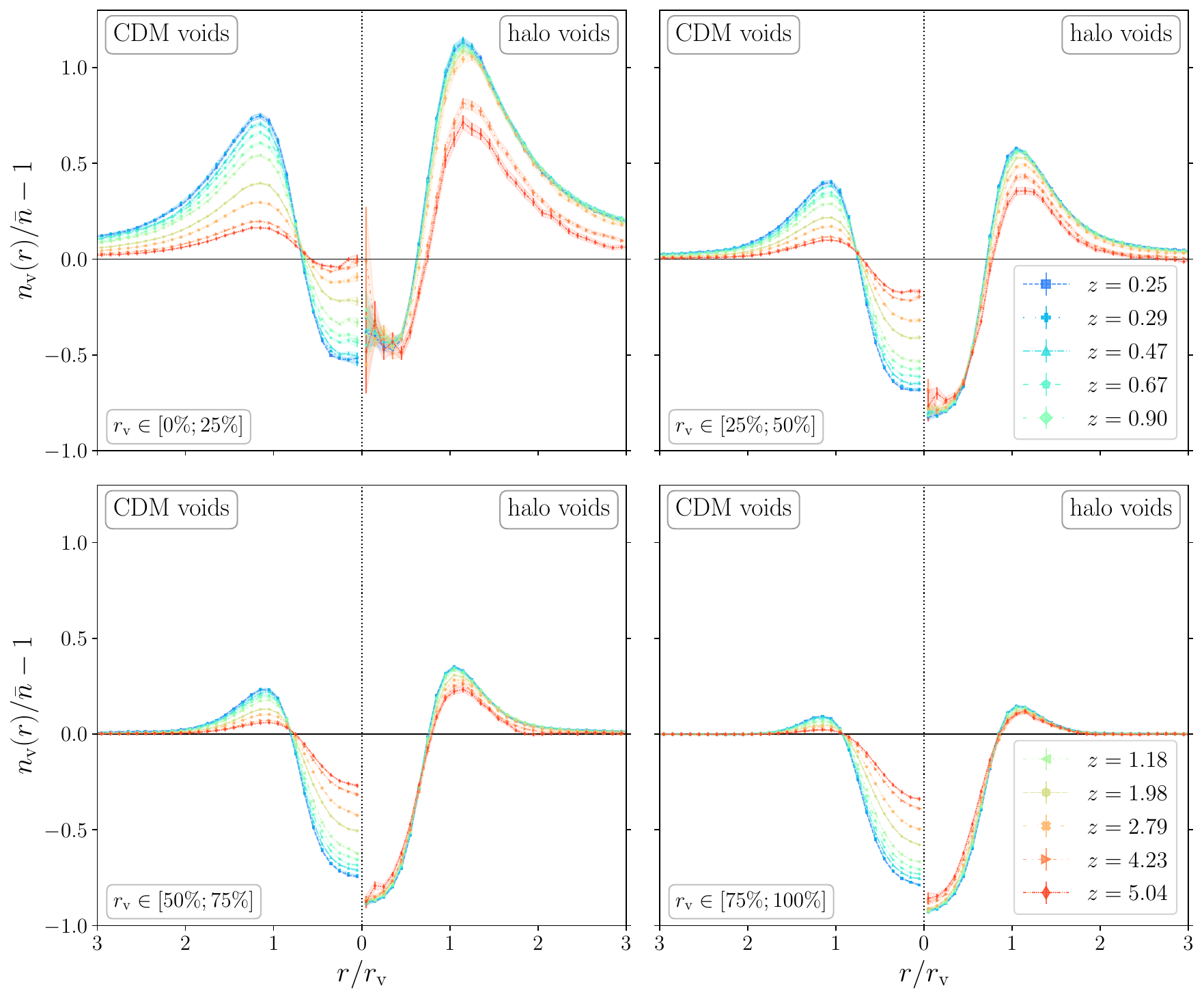}}
               \caption{Same as figure~\ref{fig_density_voids_midres_radius}, but for the \HR{} simulation and with profiles stacked in percentile bins of void radius.}
               \label{fig_density_voids_highres_relative_radius}
\end{figure}

To isolate the physical evolution of different void populations from the selection effects caused by evolving VSFs, we now re-stack their profiles in bins of relative, rather than absolute, size. We divide the total void population at each redshift into quartiles (0--25\,\%, 25--50\,\%, 50--75\,\%, and 75--100\,\%) based on their radius. This ensures that each bin contains approximately the same number of voids, with minor variations due to voids having identical radii. This method provides a more self-similar comparison across different cosmic epochs and simulation resolutions. We have also verified that using a different number of bins (e.g., quintiles) does not alter the observed trends, though it does reduce the range of scales covered in each bin, which can be advantageous for certain applications.

Figures~\ref{fig_density_voids_midres_relative_radius} and~\ref{fig_density_voids_highres_relative_radius} present the resulting profiles for the \MR{} and \HR{} simulations, respectively, with CDM voids again in the left panels and halo voids on the right. For the smallest voids, particularly the first quartile (0--25\,\%, top left), we observe two key effects. First, their large-scale environment evolves dramatically, with the density far outside the compensation wall increasing significantly over time for both CDM and halo voids. Second, particularly in \mr{}, the sparse sampling effect we noted in Figure~\ref{fig_density_voids_midres_radius} is now present at all redshifts, since this sampling artifact naturally scales with the relative void size, as both are linked to the evolving mean tracer separation. Larger CDM voids evolve similarly to those in fixed-size bins but with more pronounced compensation walls, while the denser tracer sample in \hr{} again reveals a more detailed internal substructure, resulting in higher inner densities.

For halo voids, switching to relative size bins is particularly insightful. This method is crucial because it mitigates the selection effects seen in fixed-size bins and allows for a more consistent tracking of specific void types. With this clearer view, the halo density profiles now follow the trend expected from the evolution of the underlying CDM: the compensation walls grow with time, a trend most evident in the smaller void quartiles. For the largest voids (75--100\,\% quartile), the comoving profiles are remarkably stable, showing only minor subsequent evolution at low redshifts compared to other bins, with almost perfect overlap within errors. As in the fixed-size bins, this stability in void interiors suggests a near-equilibrium between the formation of new halos within voids and the outflow of existing ones.

\subsection{Discussion and velocity profiles \label{subsec:discussion}}

The preceding analysis highlights a crucial methodological point: comparing voids of a fixed comoving size across different cosmic epochs can be misleading. As we have shown, the apparent `inverse' evolution of halo void profiles is a selection effect caused by the evolving void size function and amplified by the halo bias. Switching to a relative framework, such as quartile bins, mitigates this effect and provides a clearer view of the underlying physical evolution. The idea that a void's properties are more fundamentally tied to its rank within its contemporary population than to its absolute size is reinforced by the self-similarity observed between the \MR{} and \HR{} simulations. In this picture, voids undergo the same evolution at different cosmic epochs, with the timing dependent on the relative tracer density and the chosen minimum halo mass. The necessity of this relative framework for halo voids is highlighted by the simpler evolution of the underlying CDM density, which becomes progressively emptier without such strong selection effects.

With these methodological effects accounted for, a consistent physical picture of halo void evolution emerges. At late times ($z \lesssim 1$), the halo void population stabilizes, a phenomenon characterized by a strong alignment of their comoving density profiles, size functions, and core density distributions. Such a stabilization signals a fundamental shift: the evolution of voids is no longer dominated by the dynamic formation of halos, but is instead primarily driven by the passive expansion of the Universe. This transition occurs around the cosmic epoch where dark energy begins to significantly influence cosmic expansion. While dark energy dominates the global expansion at lower redshifts, underdense regions like voids are expected to become dominated by its influence earlier, suggesting a possible link between the accelerated expansion and the structural stabilization of voids. This large-scale stability is complemented by an internal dynamic equilibrium, where the formation of new halos within voids almost perfectly balances the outflow of existing ones.

While not depicted in this work, our analysis of velocity profiles supports these findings. Both CDM and halo voids exhibit an evolution in their dynamics consistent with their density evolution. Specifically, velocities within voids increase over time due to gravitational attraction to void boundaries, with a slight decrease at low redshifts. This `turning point' is present in all fixed-size bins presented here. The profiles of halo voids show a different trend outside their boundaries: they transition from a strong net inflow to the boundaries at high redshifts to much smaller velocities at low redshifts. In relative size bins, the evolution of velocities inside the voids shows a similar trend, while velocities outside voids show a continuous growth in their absolute values. We also confirm that halos and the underlying CDM move together around halo voids, with only minor deviations at high redshift, likely due to the sparser halo population. This reaffirms that both tracers are governed by the same large-scale cosmic flow across time~\cite{Schuster2023}.

To ensure the reliability of our findings, we additionally tested their sensitivity to the tracer selection. Our results for halo voids remain consistent whether subhalos or `true' halos (see Section~\ref{sec:Magneticum}) are used for the void identification and for measuring the density profiles. Similarly, we observe the same trends when using baryons as tracers instead of CDM. This evolution, observed across multiple tracers, provides strong support for our conclusions regarding the fundamental nature of void evolution.

\section{Linear structure growth predictions \label{sec:growth_factor}}

In Section~\ref{sec:density_profiles}, we observed that void centers become emptier as matter accumulates around their compensation walls over time. In this section, we test how well this observed evolution can be predicted by linear theory. For these predictions, we model the growth of density perturbations using the linear growth factor, $D_{+} (a)$, which for a $\Lambda$CDM cosmology is given as a function of the scale factor $a = 1 / (1+z)$ by~\cite{Dodelson2020}:

\begin{equation}
    D_{+} (a) = \frac{5 \, \Omega_\matter}{2} \, \frac{H(a)}{H_0} \,  \int_0^a \frac{ \text{d} \Tilde{a} }{ \left( \Tilde{a} \, H(\Tilde{a}) / H_0  \right)^3 } \,.
\label{eq:growth_factor_definition}
\end{equation}

Using the normalized growth factor, $\hat{D}(z, z^\star)$ ($\hat{D} = 1 $ at current time $z = 0$), we can predict the density contrast $\delta$ (i.e. our density profiles) at any redshift $z$ from a known density profile at a reference redshift $z^\star$ with the simple relation:

\begin{equation}
    \delta \left( \bm{x}, z \right)  \equiv  \rho \left( \bm{x}, z \right) / \bar{\rho} \left( z \right)  -1 = \hat{D}(z, z^\star) \, \left[ \rho \left( \bm{x}, z^\star \right) / \bar{\rho} \left( z^\star \right)  -1 \right] \,.
\label{eq:growth_factor_evolution}
\end{equation}

For our analysis, we establish the density profile at the lowest available redshift $z_\mathrm{min}$ as our baseline. We then use Equation~\ref{eq:growth_factor_evolution} to make a `backward' prediction for the profiles at all higher redshifts and compare these predictions with our measured profiles from the simulations. We find that this backward prediction is more robust than predicting forward from a high-redshift baseline. The latter approach is highly sensitive to small initial fluctuations that can lead to large deviations and artificially overestimated errors at late times, as these errors scale with $\hat{D}$. This robustness is particularly advantageous for upcoming surveys, which will provide highly constrained baselines thanks to dense tracer populations at low redshift. In combination with weak lensing data, these predictions for high-redshift voids can provide reliable information for the Alcock-Paczynski test. A trade-off of our backward method, however, is that it slightly underestimates the errors of the predictions. This occurs because the propagated error of the baseline profile becomes smaller at higher redshifts due to multiplication with $\hat{D}$. Consequently, any plots showcasing the difference between the measured data and our predictions will present a conservative estimate of the agreement. A small deviation that might, in reality, be consistent with zero (within the true error bars) could appear significant due to the reduced error bars. This means our plots depict the worst-case scenario of agreement, making small inconsistencies between the model and the data appear more pronounced.

\subsection{Evolving void populations \label{subsec:growth_factor_evolving_voids}}

\begin{figure}[t!]
               \centering \resizebox{\hsize}{!}{
                               \includegraphics[trim=0 5 0 5, clip]{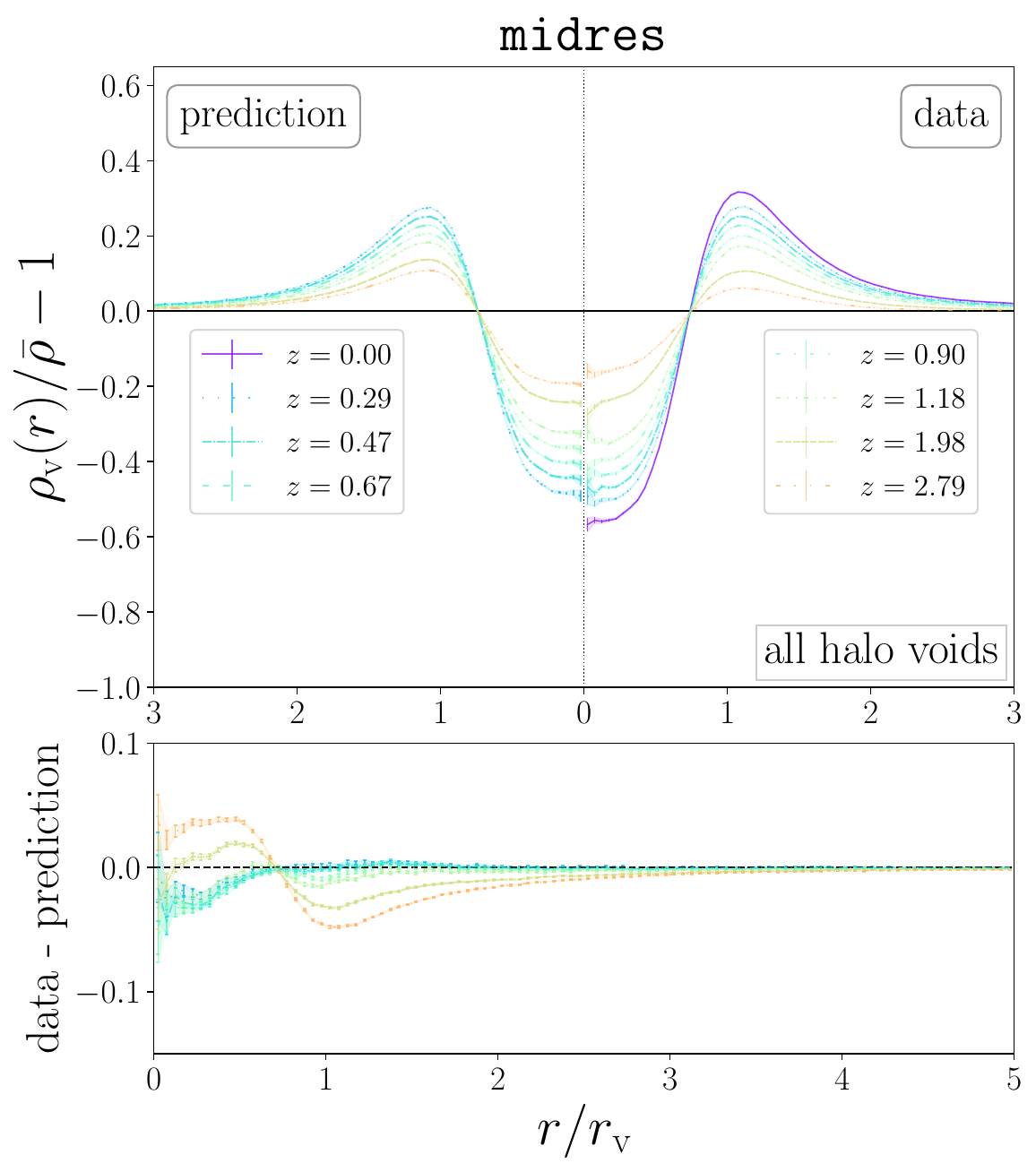}
                               \includegraphics[trim=0 5 0 5, clip]{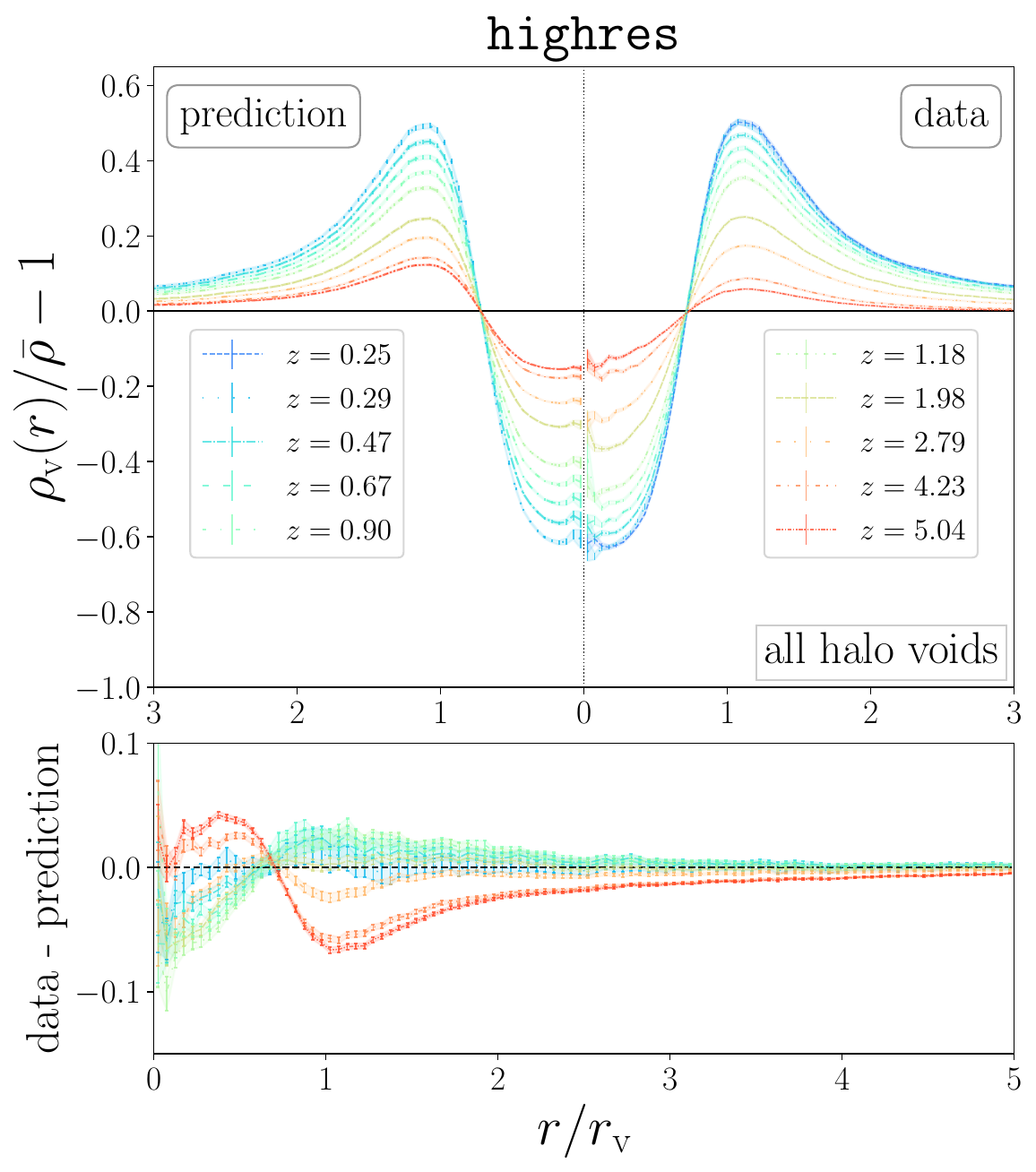}}
               \caption{Comparison between measured CDM densities (`data') around halo voids, generated by stacking all voids of each redshift into a single bin, and the corresponding linear growth factor predictions (`prediction'). Results are presented for \MR{} (left column) and \HR{} (right column). In each column, the top panel displays the prediction (left) and the data (right), while the bottom panel shows the absolute difference between them. The profiles measured in the simulations are now calculated in shells of width $0.05 \times r_\void$.}
               \label{fig_growth_factor_all_voids}
\end{figure}

\begin{figure}[t!]
               \centering \resizebox{\hsize}{!}{
                               \includegraphics[trim=0 5 0 5, clip]{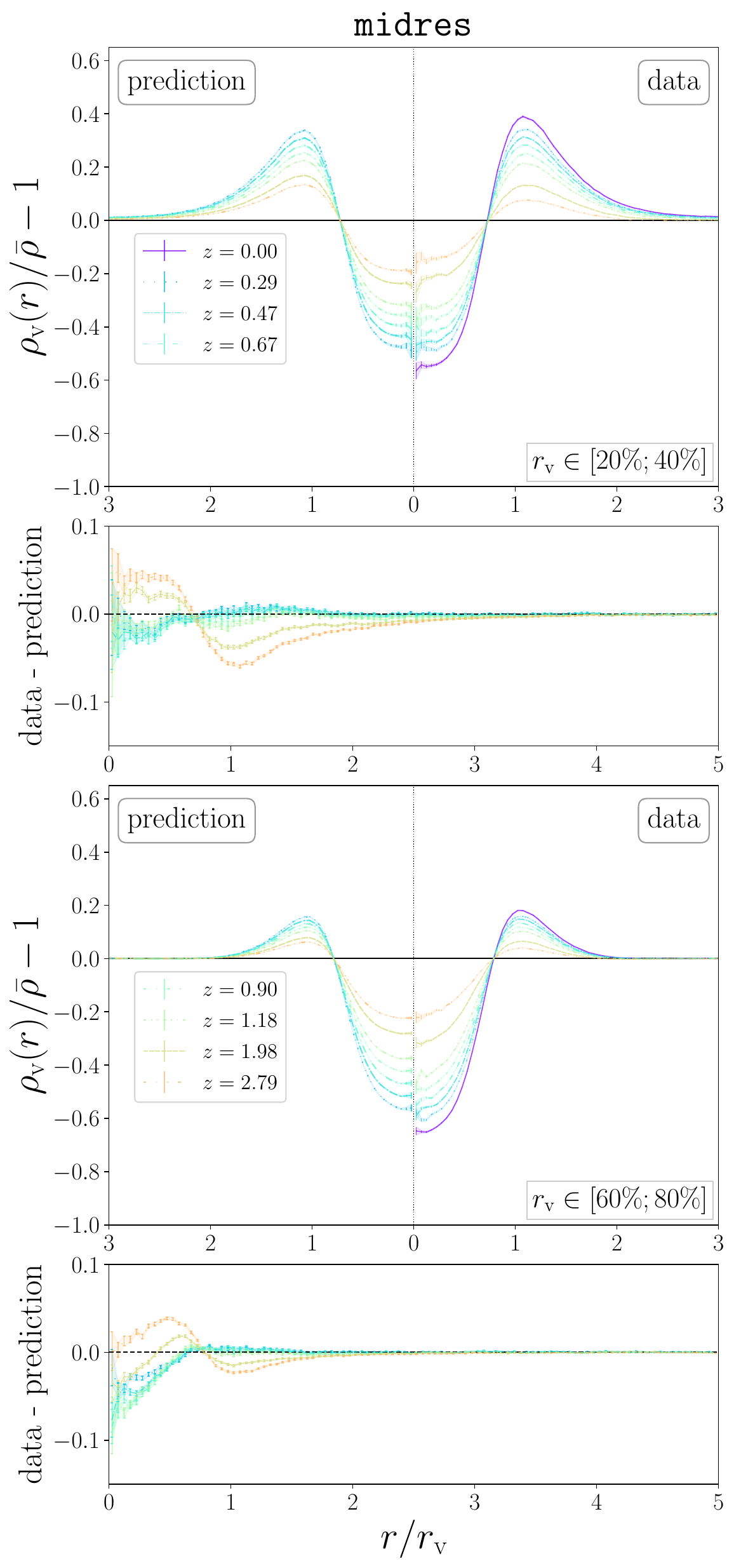}

                               \includegraphics[trim=0 5 0 5, clip]{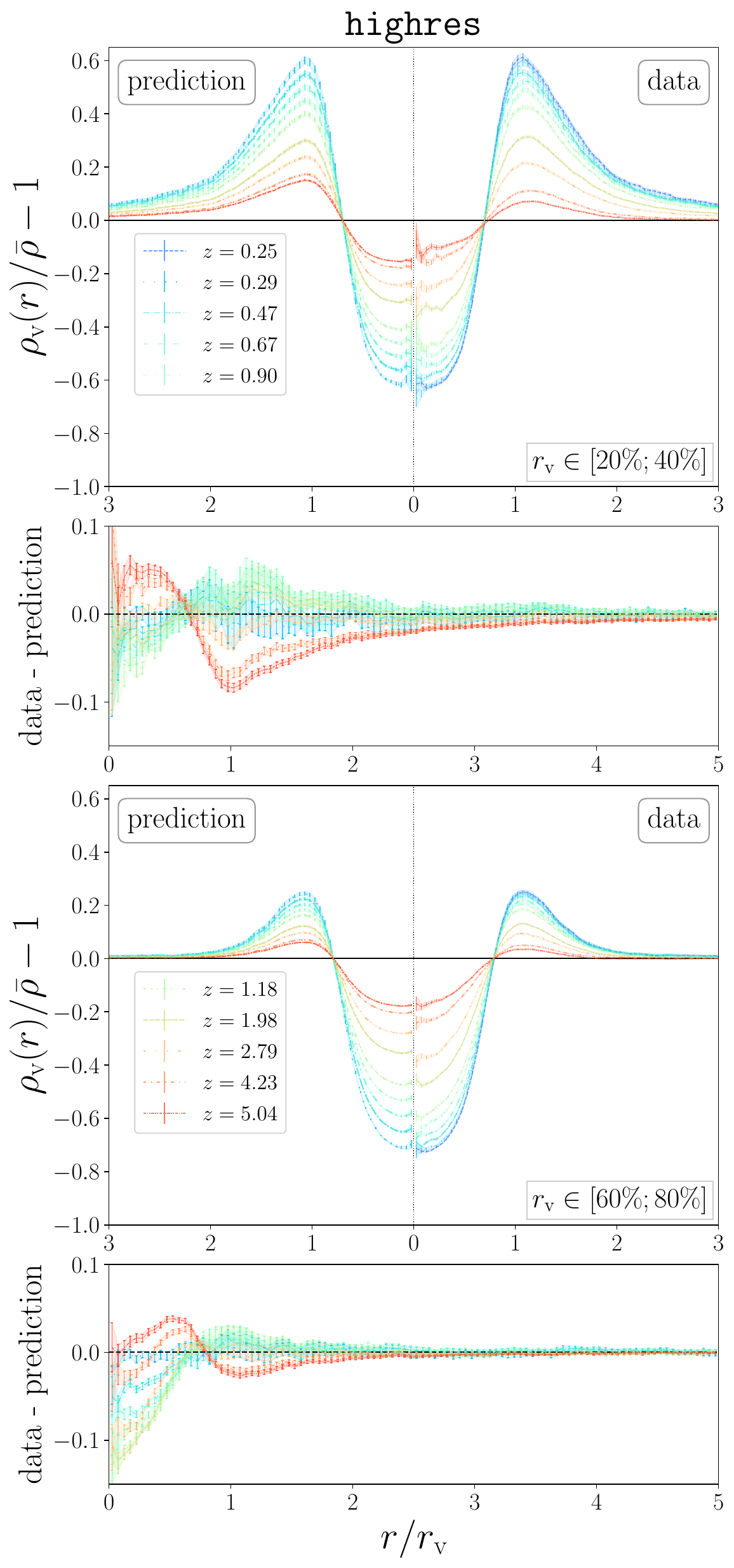}}
               \caption{Same as figure~\ref{fig_growth_factor_all_voids}, but for voids stacked in two different percentile radius bins: $20-40\,\%$ (top) and $60-80\,\%$ (bottom). This figure demonstrates that while linear growth theory is highly accurate, it deviates at the extremes, showing signatures of non-linear growth around small voids and suppressed growth in the largest voids.}
               \label{fig_growth_factor_quintiles} 
\end{figure}

Figures~\ref{fig_growth_factor_all_voids} and~\ref{fig_growth_factor_quintiles} present the predictions (left panels) and data (right panels) for CDM density profiles around halo voids in \MR{} and \HR{} (again for a denser CDM sample). To capture the coevolution of CDM and halos, voids are identified at the same redshift for which the profiles are measured. Since the growth factor itself is scale-independent, we first present the density profiles for stacks using all available voids of a given redshift in Figure~\ref{fig_growth_factor_all_voids}. This allows us to cover a wide range of scales simultaneously (see Figure~\ref{fig_void_size_functions} for the respective size distributions).

Over a wide range of cosmic history, the measured profiles accurately follow our linear theory predictions. During this period, deviations are found only in void interiors, and are of the order of $\lvert \Delta \rho /  \bar{\rho} \rvert \lesssim 0.05$ at most. The redshift range of these accurate predictions is even slightly larger than the range where many void statistics align. Deviations only become more significant at higher redshifts ($z > 1.18$ in \mr{} and $ z > 1.98$ in \hr{}), where we find a general overestimation of the compensation wall and slight underestimation of interior densities. This is consistent between both simulations, with the onset of stronger deviations once more being resolution dependent.

To test for scale-dependent effects, we now focus on specific void populations, depicting predictions for the 20--40\,\% (top) and 60--80\,\% (bottom) quintiles in Figure~\ref{fig_growth_factor_quintiles}. The results reveal two distinct regimes where the linear model deviates from the simulation data, each pointing to a different physical process. Analysis of the other quintiles supports these findings, revealing that the observed trends are even more pronounced in the most extreme populations (0--20\,\% and 80-100\,\%).

In the smallest voids, the primary discrepancy is an overprediction of the compensation wall height at high redshifts, which is a signature of non-linear evolution. The small-scale compensation walls of these voids undergo strong non-linear gravitational growth, becoming taller than the simple linear growth model predicts. Since our backwards prediction is based on the final state (higher walls), it incorrectly infers a higher initial wall to compensate for the slower growth, leading to the observed overestimation.

In the largest voids, a different and more profound effect emerges. While the compensation walls are predicted with high accuracy, with their densities only slightly overestimated, the growth model consistently overestimates the interior densities at intermediate redshifts. This overestimation from a low redshift baseline implies that matter is evacuating these large voids more slowly than linear theory suggests to reach the final densities --- a clear sign of suppressed late-time structure growth. This highlights a different type of non-linear growth. The physical lower bound of $\delta = -1$  ($\rho = 0$) necessitates a departure from linear theory, which would otherwise predict non-physical negative densities in the most underdense regions at some point in time. A compelling physical explanation for this effect could be the local impact of dark energy. As the most underdense regions in the Universe, these large voids are the first places to be dominated by the influence of dark energy due to their lower local matter density, leading to an earlier onset of cosmic acceleration that slows down the growth of structure within them~\cite{Williams2025}. This could also provide a physical mechanism for the `freezing' of void structures observed in our analysis of the halo number densities.

Our analysis reveals that the success of the linear model depends critically on the chosen void population. While the model works remarkably well for the matter field around halo voids, its predictive power diminishes for more dynamic populations like CDM-identified voids. This highlights a key conclusion: the model is most powerful when used with a stable, more passively evolving tracer population. This is best achieved with late-time halo voids analyzed in a relative framework, which effectively isolates the large-scale growth of the underlying matter field. Our tests show this conclusion is robust, as the model works equally well when using baryons instead of CDM.

\subsection{Impact of resolution and scale on structure growth in fixed voids \label{subsec:growth_factor_fixed_voids}}

\begin{figure}[t!]
               \centering \resizebox{\hsize}{!}{
                               \includegraphics[trim=0 5 0 5, clip]{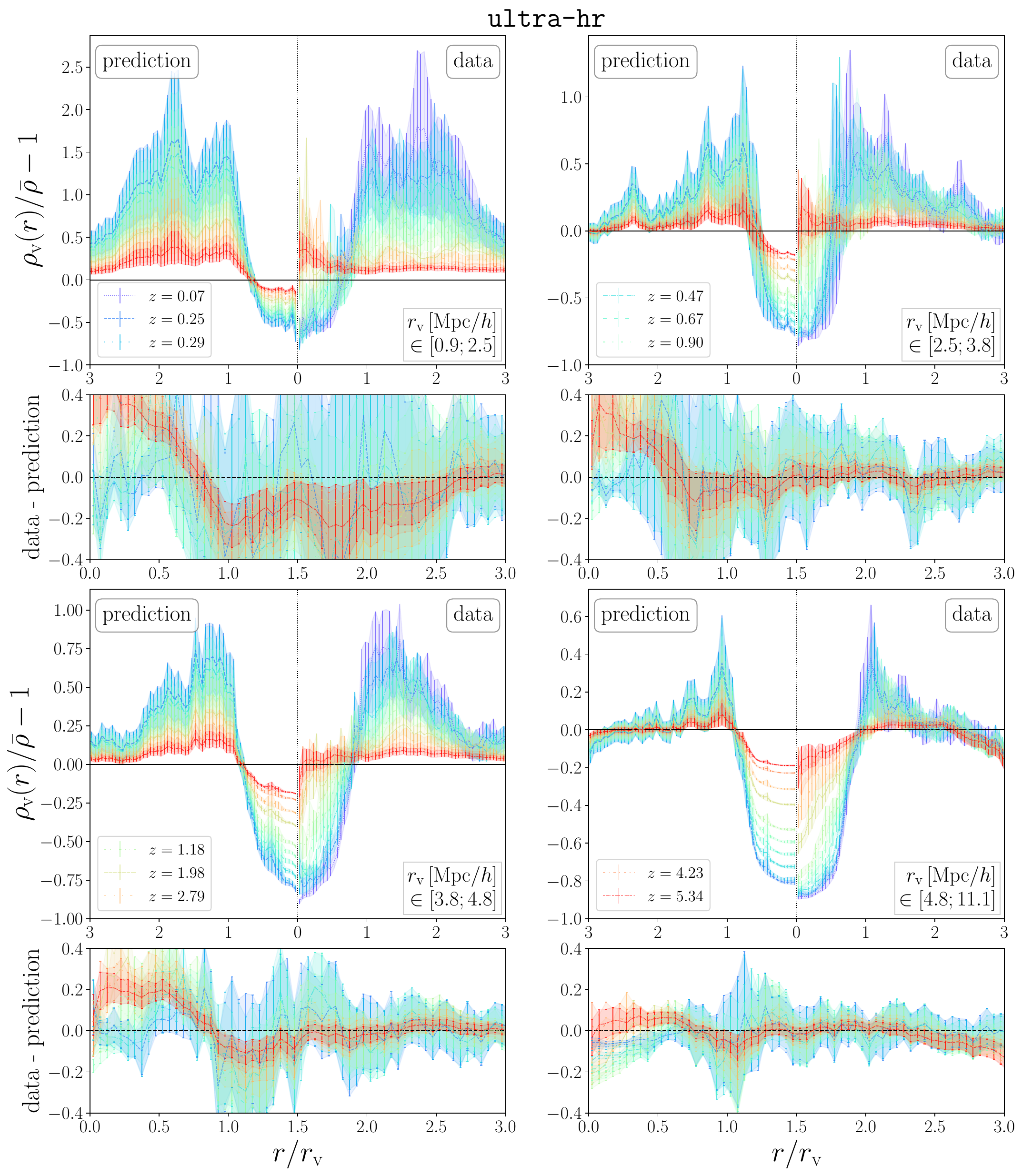}}
               \caption{Comparison between measured CDM density profiles (`data') and linear growth factor predictions (`prediction') for halo voids in the \UHR{} simulation. Voids are identified in a halo sample with a mass cut of $2 \times 10^9 \Msun$ at $z_\mathrm{min} = 0.07$, and the evolution of the surrounding CDM density is tracked over time. Each panel shows the comparison for a different quartile of the void radius distribution, with the corresponding physical sizes indicated in each panel.}
               \label{fig_growth_factor_zmin_uhr}
\end{figure}

\begin{figure}[!htbp]
    \centering
    \includegraphics[width=1.0\textwidth, trim=0 43 0 5, clip]{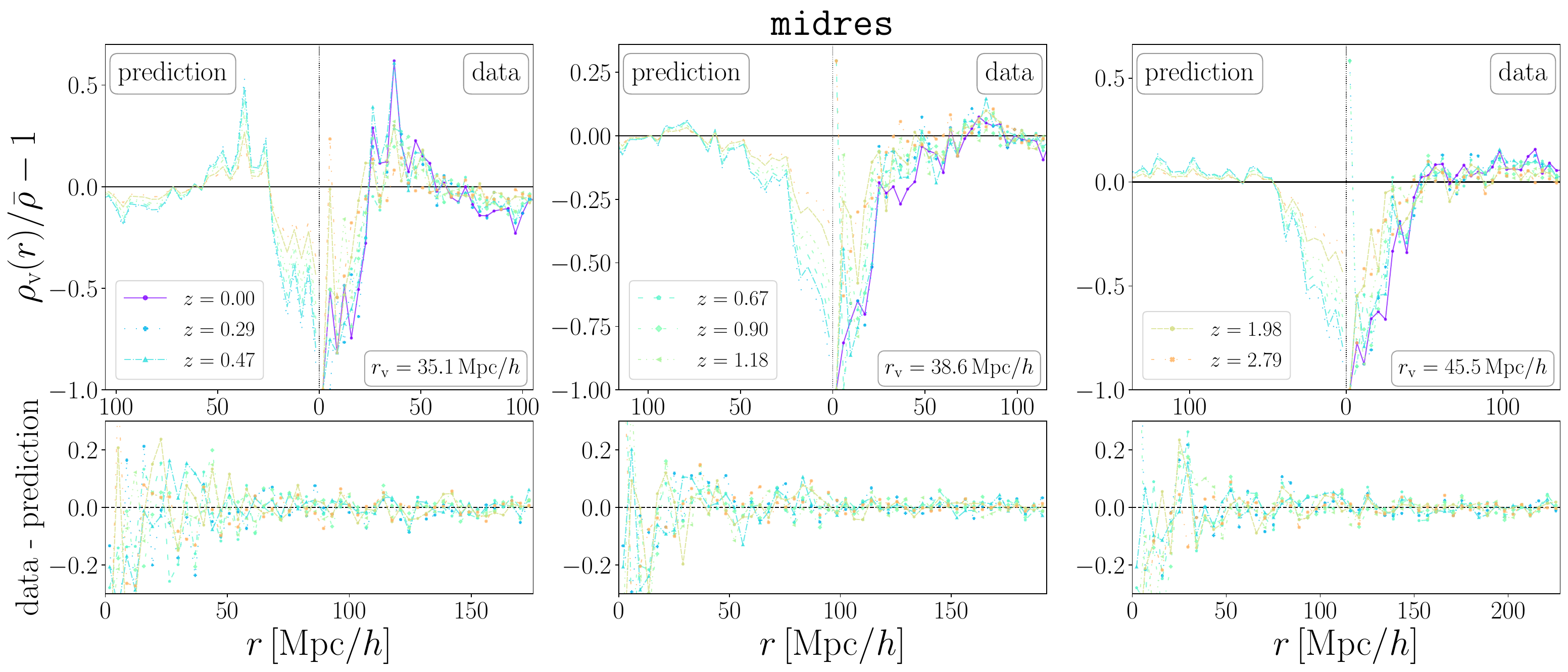}
    \includegraphics[width=1.0\textwidth, trim=0 43 0 5, clip ]{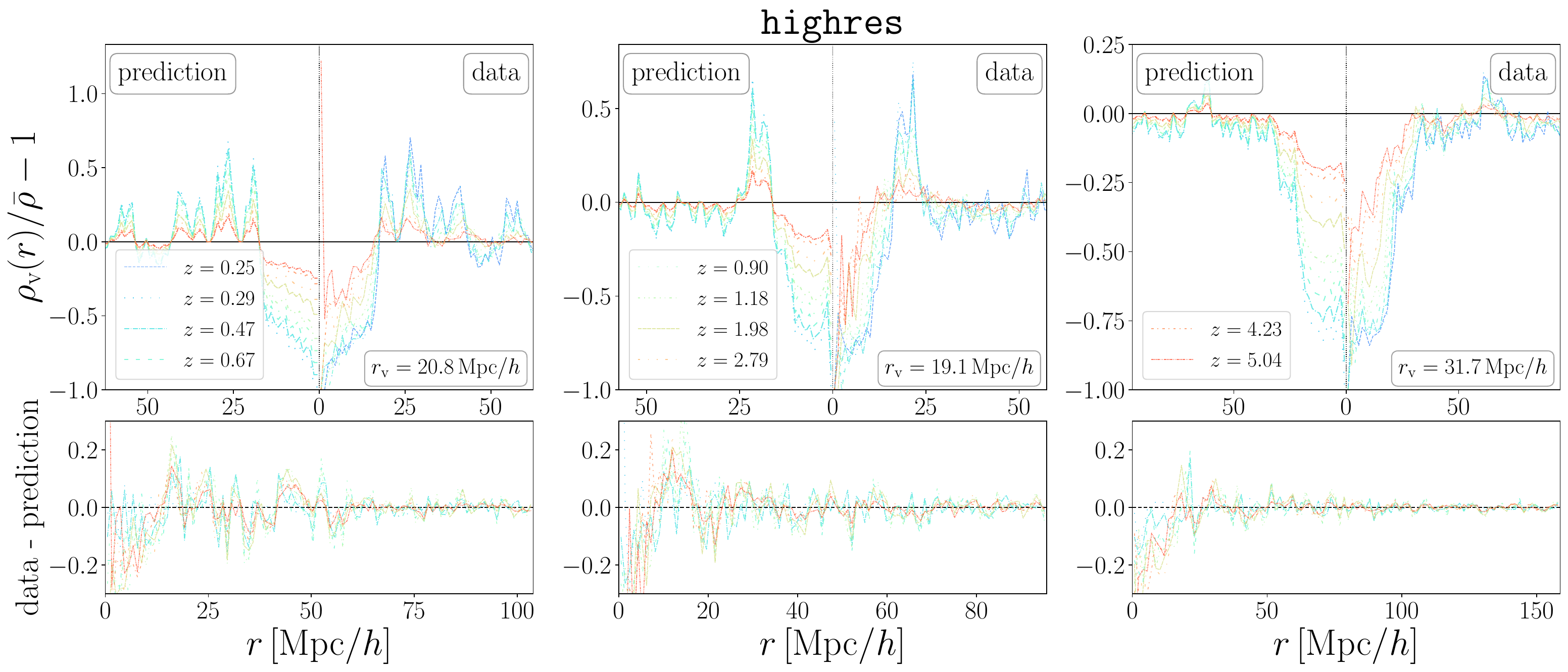}
    \includegraphics[width=1.0\textwidth, trim=0 5 0 5, clip]{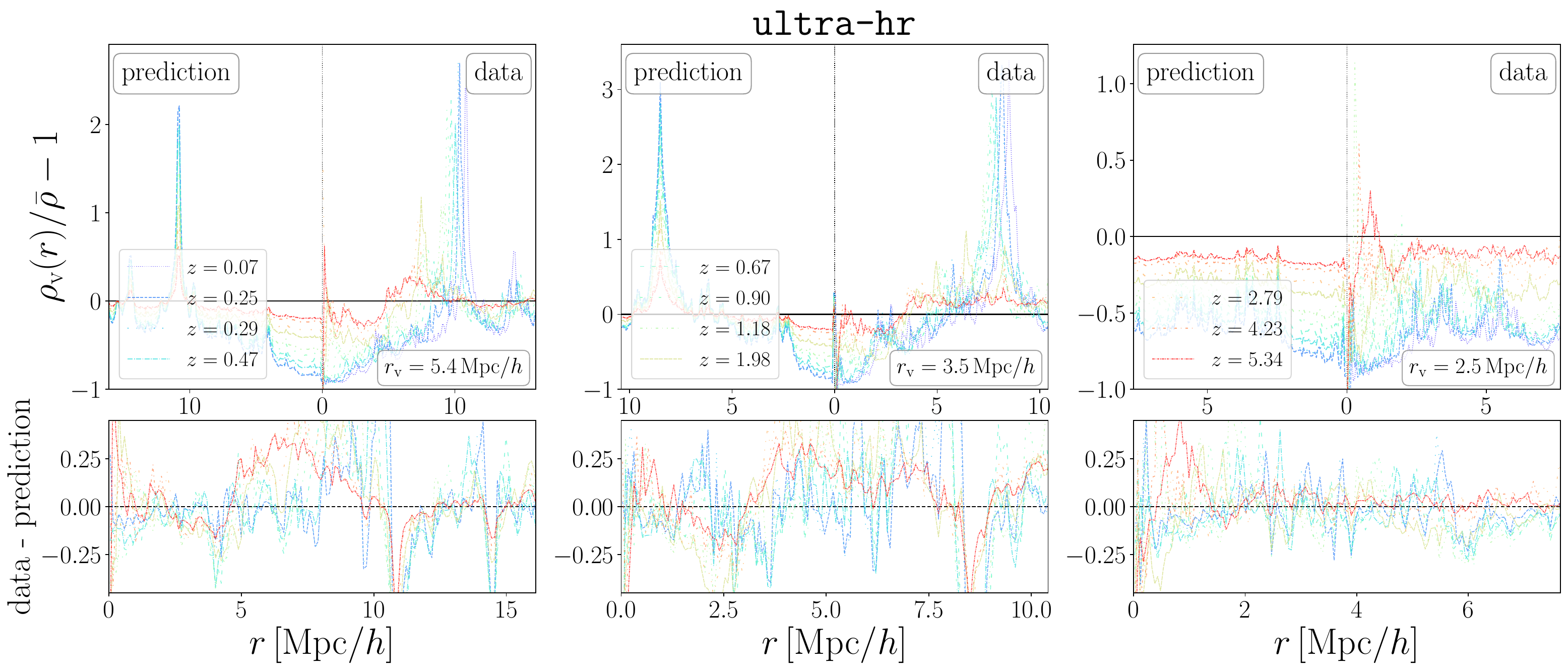}
    \caption{Comparison of linear growth factor predictions (left side of each panel) and measured CDM density evolution (right side) around selected individual halo voids identified at $z_\mathrm{min}$. Each row showcases three individual voids from a different simulation: \MR{} (top), \HR{} (middle), and \UHR{} (bottom). The density profiles are calculated using different spherical shell widths for each simulation: $0.1 \times r_\void$ (\mr{}), $0.05 \times r_\void$ (\hr{}), and $0.02 \times r_\void$ (\uhr{}), with a higher-density CDM subsample used for the \hr{} and \uhr{} calculations. Contrary to previous figures, the radial axis is shown in comoving distance ($\Mpch$) rather than in units of the void radius ($r / r_\void $).}
    \label{fig_growth_factor_individual_voids}
\end{figure}

Because the linear growth model works best for more stable and passively evolving void populations, we now apply it to a fixed set of voids identified only at our lowest redshifts, $z_\mathrm{min}$, of each respective simulation. The predicted profiles are therefore identical to those in the previous section, and only the measured CDM densities around those fixed voids at higher redshifts are different. This allows us to track the evolution of the matter field within a static reference frame, removing the effects of a dynamically evolving void population. At the same time, this static frame also means that any large-scale motions are not accounted for, including the voids' own displacement, which can potentially lead to increased deviations between the predictions and data.

To further investigate the scale-dependence of the growth model, we focus on the \UHR{} simulation (see Section~\ref{sec:Magneticum} for more details). While its small box size ($48\,\Mpch$) and limited halo void sample ($241$ voids at $z_\mathrm{min} = 0.07$ for $M_\halo \geq 2.0 \times 10^{9} \, \Msun$) restricted its previous use for the analysis of evolving void populations, these limitations are not relevant when analyzing fixed, late-time voids. Instead, the \uhr{} simulation now provides a unique opportunity to test linear theory on extremely small scales, similar to its use in previous void studies~\cite{Schuster2023,Schuster2024}. To ensure robust measurements on these scales, especially given the small number of voids, the CDM densities are calculated using an extremely dense subsample of $6 \times 10^7 $ particles, comprising approximately 31\,\% of all CDM tracers in \uhr{}.

We refrain from presenting results for stacked \mr{} and \hr{} voids, as their agreement with the linear theory is very similar to that seen in Section~\ref{subsec:growth_factor_evolving_voids}, with only slightly larger deviations for smaller voids, which are more sensitive to local dynamics. We therefore focus our analysis on two more illustrative cases: Figure~\ref{fig_growth_factor_zmin_uhr} presents the stacked void population in the \uhr{} simulation, while Figure~\ref{fig_growth_factor_individual_voids} depicts a selection of individual voids from all three resolutions.

In Figure~\ref{fig_growth_factor_zmin_uhr}, the \uhr{} voids are stacked in quartiles based on their size, where the given bin edges now present their comoving radii. As anticipated from the small void numbers, the measured profiles exhibit large statistical errors. For the three smallest quartiles ($r_\void \lesssim 4.8\,\Mpch$), the linear growth prediction fails significantly at high redshifts. Notably, the central densities at $ z = 5.34$ around these late-time voids are measured to be near or even above the cosmic mean. Such a state cannot be linearly extrapolated from their underdense nature at $z_\mathrm{min}$. This discrepancy is a direct consequence of the complex, non-linear dynamics governing small-scale structures. Voids of this size are subject to significant displacement over cosmic time, meaning that the initial underdensity that seeded their formation was likely located far from their late-time center. Moreover, the formation of some of these extremely small voids ($r_\void \lesssim 5 \,\Mpch$) may itself be a late-time phenomenon, determined only after the surrounding matter field had significantly clustered and the first galaxies formed. These results reinforce the conclusion from other studies that linear theory is less reliable for describing the evolution of voids with radii smaller than $5 \, \Mpch$~\cite{Stopyra2021}.

In contrast, the largest void quartile (with radii from $4.8$ to $11.1\,\Mpch$) exhibits the expected deep central underdensities at high redshift. While the overall alignment between the linear predictions and measurements is very good, with statistical errors on the order of $\lvert \Delta \rho  /  \bar{\rho} \rvert \pm 0.15 $, we also confirm the trend of suppressed growth for large voids identified in the previous Section~\ref{subsec:growth_factor_evolving_voids}. The model again consistently overestimates the interior densities at low and intermediate redshifts. This indicates that at these redshifts matter evacuates void interiors more slowly than linear theory suggests, a deviation that warrants further modeling. The slight dip in the measured profiles at large radii is a known artifact of the small simulation box size. The finite volume limits the number of unique particles available for stacking, which artificially suppresses the profile and prevents it from returning to the cosmic mean density.

To complement the analysis of stacked late-time voids, Figure~\ref{fig_growth_factor_individual_voids} illustrates the CDM density evolution around nine representative voids, with three selected from each simulation. These examples were chosen as typical cases to schematically demonstrate the fundamental robustness of the growth model at the level of individual structures, beyond its success in statistical averages.

A recurring pattern is visible in the residuals (lower panels), which show the absolute difference between data and prediction. Crucially, without statistical error bars, these plots represent the absolute worst-case scenario for the model's performance. In these residuals, an overestimation of density at one radius is often adjacent to an underestimation at a neighboring radius, creating an almost oscillating pattern. This suggests that many small-scale discrepancies are caused by localized matter displacement, such as halos moving between adjacent shells, a small-scale dynamic not captured by the linear approximation.

The model's performance varies predictably with void characteristics. Linear theory breaks down most significantly in two scenarios. First, for the smallest voids in each simulation, where local dynamics and resolution effects degrade the predictive accuracy. This effect is evident in the larger y-axis range of the difference panel in \uhr{}, where absolute density deviations are most significant. Second, for voids that develop highly clustered compensation walls by $z_\mathrm{min}$, the model consistently overestimates the density within those walls at earlier times, as it cannot capture extreme non-linear collapse, an effect we already observed in stacked profiles. Conversely, the model performs remarkably well for density depressions already present at high redshift. Such underdensities evolve more passively as matter evacuates their centers, a behavior the linear model approximates accurately, except for minor local clustering (e.g., the rightmost void in \UHR{}).

\section{Conclusion\label{sec:conclusion}}

This paper explored the formation and evolution of cosmic voids and their properties from redshift $z = 5.04$ to the present day. By analyzing hydrodynamical simulations from the \Mag{} suite, we were able to study void evolution across a wide range of scales and resolutions. Defining voids with both dark matter particles and biased halos allowed us to uncover their unique evolutionary dynamics and quantify the impact of tracer selection. This approach is crucial for understanding upcoming observational datasets and for testing the limits of linear growth predictions of these cosmic environments. Our primary findings are as follows:

\begin{itemize}

    \item \textbf{Divergent evolution of void populations:} We find that CDM- and halo-defined voids follow starkly different evolutionary tracks in both their numbers and sizes. The number of CDM voids decreases over time as they hierarchically merge and grow larger, which is reflected in their void size functions shifting towards larger radii. In contrast, halo voids evolve through fragmentation. As new halos form, they continuously split the vast, sparse voids of the early Universe, leading to a greater number of smaller voids at low redshift (Figures~\ref{fig_tracer_void_numbers} and~\ref{fig_void_size_functions}).

    \item \textbf{Contrasting evolution of global void properties:} While their ellipticities evolve minimally over time, the internal densities of voids show clear differences. The core densities of CDM voids become progressively lower as matter evacuates their centers. Halo voids, however, maintain a remarkably stable core density distribution, suggesting an equilibrium between the outflow of existing halos and the formation of new ones within voids. While the overall distribution of void compensations broadens with time, the medians for both types remain exceptionally stable, a consequence of being established early and largely predetermined by the initial cosmic density field (Figure~\ref{fig_void_property_evolution}).

    \item \textbf{The standard growth of CDM voids:} The density profiles of CDM voids evolve as expected under the influence of gravity and structure formation. Over time, their interiors become progressively emptier as matter streams outwards, causing their surrounding compensation walls to grow in height and density. For smaller voids in particular, the large-scale environment far beyond these walls also evolves, growing significantly denser as structures evolve (Figures~\ref{fig_density_voids_midres_radius}, \ref{fig_density_voids_highres_radius}, \ref{fig_density_voids_midres_relative_radius}, and~\ref{fig_density_voids_highres_relative_radius}).

    \item \textbf{Late-time stabilization of halo voids:} The halo void population stabilizes at low redshifts ($z \lesssim 1$). This is evident in their density profiles, which align in comoving coordinates, signaling a shift where cosmic expansion becomes the dominant driver of their evolution, rather than active halo formation and halo dynamics. Crucially, the underlying CDM density around these same voids continues to evolve as expected, with their interiors deepening and walls growing even after the halo population has stabilized (Figures~\ref{fig_density_halo_voids_CDM_midres_radius}, \ref{fig_density_halo_voids_CDM_highres_radius}, \ref{fig_density_voids_midres_relative_radius}, and~\ref{fig_density_voids_highres_relative_radius}).

    \item \textbf{The importance of a relative size framework:} We demonstrate that comparing halo voids of a fixed comoving size across cosmic time introduces a significant selection effect due to mixing populations, leading to an apparent `inverse' evolution of their number density profiles. This effect, which is additionally amplified by the evolution of halo bias, is a methodological artifact, not a physical process. By switching to a relative framework that bins voids by their percentiles in size, this effect is mitigated, restoring the expected underlying physical evolution and providing a robust method for analyzing observational data (Figures~\ref{fig_density_halo_voids_CDM_midres_radius}, \ref{fig_density_halo_voids_CDM_highres_radius}, \ref{fig_density_voids_midres_relative_radius}, and~\ref{fig_density_voids_highres_relative_radius}).

    \item \textbf{Self-similarity across scales:} Our findings across all sections reveal a remarkable self-similarity in void evolution between the \MR{} and \HR{} simulations. The different resolutions and halo mass cuts effectively probe the same physical evolutionary stages --- from fragmentation to stabilization --- simply shifted to different scales and cosmic epochs depending on the properties of the underlying tracers. This confirms that the internal structure of voids is self-similar and scales with the mean tracer separation, a trend observed across the various statistics and profiles analyzed in this work.

    \item \textbf{Success of linear growth predictions:}  The simple linear growth factor model proves remarkably successful at predicting the evolution of the CDM density around stable, late-time halo voids. For stacked void profiles, deviations are typically less than $\lvert \Delta \rho /  \bar{\rho} \rvert \lesssim 0.05$ over a wide range of cosmic history. Furthermore, the model remains fundamentally robust even for tracking the density evolution around individual voids fixed at $z_\mathrm{min}$, confirming its success beyond statistical averages. This overall agreement confirms that void interiors are pristine laboratories where the large-scale dynamics of matter are well-described by linear theory (Figures~\ref{fig_growth_factor_all_voids}, \ref{fig_growth_factor_quintiles}, \ref{fig_growth_factor_zmin_uhr}, and~\ref{fig_growth_factor_individual_voids}).

    \item \textbf{Scale-dependent deviations from linearity:} We identify distinct regimes where linear growth predictions deviate more significantly from the data. First, for small voids in general, the compensation walls grow stronger than predicted, a clear signature of non-linear gravitational growth. Second, for the smallest voids ($r_\void \lesssim 5 \,\Mpch $)  analyzed in a fixed location at $z_\mathrm{min}$, the model fails to account for their complex formation history, in which their high-redshift progenitors were not necessarily underdense yet or
    were displaced from the final void center (Figure~\ref{fig_growth_factor_zmin_uhr}). Finally, for the largest voids, we observe evidence of suppressed late-time structure growth, where matter evacuates their interiors more slowly than linear theory suggests, a non-linear trend possibly explained by the early local influence of dark energy (Figure~\ref{fig_growth_factor_quintiles}).
    
\end{itemize}

Our work shows that once a complete sample of low-mass halos (the likely counterparts to low-brightness galaxies) is observed, the halo void population effectively `freezes' in comoving coordinates at late times. This stability transforms them into pristine laboratories for studying the nature of dark energy, as they become passive tracers of cosmic expansion. With recent hints of a dynamic dark energy from DESI~\cite{DESI_DR1_constraints_bao, DESI_DR1_constraints_full_shape, DESI_DR2_constraints_bao}, this finding warrants detailed studies into the impact of various dark energy models on void evolution. The suppressed late-time growth observed in the largest voids may provide a physical picture of this process, a phenomenon that warrants further investigation to enhance the power of voids as cosmological probes. With a wealth of data from ongoing DESI observations and the datasets expected from Euclid~\cite{Hamaus2022, Contarini2022, Bonici2023, Radinovic2023}, the study of voids to probe the nature of dark energy is poised to become a highly relevant frontier in the coming years.

Moreover, our paper establishes a robust relative framework essential for extracting clean cosmological signals. This methodology should be adopted in future analyses to cleanly track void evolution, as it successfully disentangles true physical growth from tracer-related selection effects. Doing so will be highly relevant for cosmological inference via the Alcock-Paczynski effect and redshift-space distortions (RSD)~\cite{Hamaus2016,Hamaus2020,Correa2021b,Hamaus2022}. Beyond illuminating the evolution of voids, this relative size framework could also be valuable for analyzing scenarios where voids show inverse trends, such as the effects of massive neutrinos~\cite{Kreisch2019,Schuster2019}. In addition, the close agreement between CDM density evolution and linear predictions can enable powerful new void weak lensing tests across a wide range of redshifts~\cite{Melchior2014,Clampitt2015,SanchezC2017,Davies2018,Fang2019,Davies2021,Jeffrey2021}. Beyond dark energy, this work opens other avenues to probe fundamental physics. The distinct evolutionary paths of dark matter and halos offer a new arena to test gravity and constrain the nature of dark matter. Furthermore, the stability of void interiors makes them ideal laboratories to determine the neutrino mass scale and to search for subtle, preserved imprints from the early Universe, such as signatures of inflation or primordial non-Gaussianity. Since this work focused on isolated voids, we briefly comment on our expectations for merged voids. Here, a potential systematic effect could arise, as the internal sub-structures increase the density of larger voids but not smaller ones, which are unaffected by merging (see~\cite{Schuster2023}). This size-dependent effect would introduce a redshift-dependent scale that could mask or obscure our reported results.

Crucially, by identifying the precise epochs and scales at which the linear growth model holds, this work provides essential guidance for void analyses in ongoing and future cosmological surveys. The combination of their linear dynamics across vast scales~\cite{Schuster2023}, the relative insignificance of baryonic physics~\cite{Schuster2024}, and their overall pristine evolution makes cosmic voids exquisite laboratories for cosmology. While our analysis was performed for a specific cosmology, we expect these fundamental properties to be robust for observationally relevant $\Lambda$CDM models. Variations in parameters like $\Omega_\matter$ will likely only shift the scale and time at which voids stabilize. The vast number of voids from upcoming datasets and their unique characteristics provide a clear path forward for testing the underlying physical principles of our Universe and performing cosmological inference, paving the way for a new era of void-based precision cosmology.

\begin{acknowledgments}
We thank Adrian E. Bayer, Anna Porredon, Ben Wandelt, Carlos M. Correa, Giovanni Verza, Jahmour J. Givans, Rien van de Weygaert, Sankarshana Srinivasan, Steffen Hagstotz, as well as the participants of the Voids@CPPM 2025 workshop for fruitful discussions. This research has been funded by the Deutsche Forschungsgemeinschaft (DFG, German Research Foundation) -- HA 8752/2-1 -- 669764. The authors acknowledge additional support from the Excellence Cluster ORIGINS, which is funded by the DFG under Germany's Excellence Strategy -- EXC-2094 -- 390783311. AP and NS acknowledge support from the french government under the France 2030 investment plan, as part of the Initiative d’Excellence d’Aix-Marseille Universit\'e - A*MIDEX AMX-22-CEI-03. AP acknowledges support from the European Research Council (ERC) under the European Union's Horizon programme (COSMOBEST ERC funded project, grant agreement 101078174). KD acknowledges funding for the COMPLEX project from the European Research Council (ERC) under the European Union’s Horizon 2020 research and innovation program grant agreement ERC-2019-AdG 882679. The calculations for the hydrodynamical simulations were carried out at the Leibniz Supercomputing Center (LRZ) under the project pr83li. We are especially grateful for the support by M. Petkova through the Computational Center for Particle and Astrophysics (C2PAP) and for the support by N. Hammer at LRZ when carrying out the {\it Box0} simulation within the Extreme Scale-Out Phase on the new SuperMUC Haswell extension system. We used \texttt{matplotlib}~\cite{matplotlib} for plots. Many of the computations were done with the help of \texttt{NumPy}~\cite{NumPy} and \texttt{SciPy}~\cite{SciPy}.
\end{acknowledgments}

\section*{Data availability}

The data depicted in this article, along with the code base used to calculate void profiles, is publicly available {here}\footref{note:data_url}, while additional data (such as corresponding velocity profiles and different binnings) is available upon request to the corresponding author.

\bibliography{ms}
\bibliographystyle{JHEP}

\end{document}